\documentclass[12pt]{article}
\usepackage{amsmath}
\usepackage{graphicx}
\usepackage{enumerate}
\usepackage{natbib}
\usepackage{url} 

\usepackage{bm}
\usepackage{amsfonts}
\usepackage{subfigure}

\newcommand{\blind}{0}

\newtheorem{theorem}{Theorem}
\newtheorem{condition}[theorem]{Condition}
\newtheorem{proposition}[theorem]{Proposition}

\begin{document}

\def\spacingset#1{\renewcommand{\baselinestretch}%
{#1}\small\normalsize} \spacingset{1}


\if0\blind
{
\title{Inference of Sample Complier Average Causal Effects in Completely Randomized Experiments}
\author{Zhen Zhong\\
		Yau Mathematical Science Center, Tsinghua University\\
		\\
		Per Johansson\\
		Department of Statistics, Uppsala
		University\\
		and Yau Mathematical Science Center,
		Tsinghua University\\
		\\
		Junni L. Zhang*\\
		National School of Development and Center for Statistical Science,
		\\Peking University}
  \maketitle
} \fi

\if1\blind
{
  \bigskip
  \bigskip
  \bigskip
  \begin{center}
    {\LARGE\bf Inference of Sample Complier Average Causal Effects in Completely Randomized Experiments}
\end{center}
  \medskip
} \fi

\begin{abstract}
In randomized experiments with non-compliance scholars have argued that the complier average causal effect (CACE) ought to be the main causal estimand. The literature on inference of the complier average treatment effect (CACE) has focused on inference about the population CACE. However, in general individuals in the experiments are volunteers. This means that there is a risk that individuals partaking in a given experiment differ in important ways from a population of interest. It is thus of interest to focus on the sample at hand and have easy to use and correct procedures for inference about the sample CACE. We consider a more general setting than in the previous literature and construct a confidence interval based on the Wald estimator in the form of a finite closed interval that is familiar to practitioners. Furthermore, with the access of pre-treatment covariates, we propose a new regression adjustment estimator and associated methods for constructing confidence intervals. Finite sample performance of the methods is examined through a Monte Carlo simulation and the methods are used in an application to a job training experiment.
\end{abstract}

\noindent%
\vfill

\newpage
\spacingset{1.9} 

\section{Introduction}
One requirement for the inference to an average treatment effect from a randomized experiment to be valid is that all experimental units comply to their treatment assignments. In practice, incomplete compliance to the assigned treatment is common.

One standard approach is to ignore the information on compliance behavior and to focus on the intention-to-treat (ITT) analysis.  It is argued that the ITT-effect may be most policy-relevant, since one cannot in general force people to take a treatment.  However, the ITT-effect may be misleading, for example, when a treatment with negative side effects may appear less effective simply because subjects adhere to it to a greater extent.  The international guidelines for good clinical practice \citep{dixon1999international}, for example, suggest that the analysis of negative side effects should be according to treatment received.  Yet the ``as-treated'' analysis that compares those who receive treatment with those who receive control generally yields a biased estimate of the treatment effect, because the treatment and control groups are no longer similar.

When information on treatment received is available, which often is the case, the effect for those who comply to the assigned treatment, i.e., the compliers, can be identified under certain conditions, and scholars have argued that the complier average causal effect (CACE) ought to be the main causal estimand (see e.g. \cite{McNamee_2009,Shrier_etal_2014,Shrier_etal_2017,Steele_etal_2015}).  In the literature, often the experimental units are regarded as being independently and randomly sampled from a super-population, and inference is made about the population CACE.  However, in general individuals in the experiments are volunteers which means that there is a risk that they differ in important ways from a population of interest.

This paper focuses on the finite population consisting of the experimental units, and thus on inference about the sample CACE.
As \cite{athey2017econometrics} points out, the line of research that only focuses on the sample at hand originates in the work by \cite{neyman1923} (translated to English in 1990) and \cite{neyman&etal:1935}.  Neyman was interested in estimating the average effect of treatment assigned for the sample at hand.  In this paper we extend this line of research to estimate the average effect of treatment received for the compliers in the sample at hand.

When pre-treatment covariates are not considered, we show that the Wald estimator which is typically used to estimate population CACE can also be used to estimate sample CACE. For construction of a confidence interval based on the Wald estimator, \cite{li2017general} assumes that the individual treatment effect is homogeneous among compliers. This allows them to derive four possible forms of the confidence sets for sample CACE, including an empty set, a finite closed interval, the whole real line, and a union of two infinite half-open intervals. The homogeneous assumption, or linear instrumental variable model, is overly restrictive. Our framework derives finite confidence intervals familiar to practitioner that puts no restriction on the individual treatment effects among compliers.

To improve efficiency, we also consider adjusting for pre-treatment covariates, and propose a regression adjustment estimator and associated methods for constructing confidence intervals.

The next section defines the sample CACE. Section 3 discusses the Wald estimator and methods for constructing confidence intervals based on the Wald estimator. Section 4 discusses the regression adjustment estimator and methods for constructing confidence intervals based on the regression adjustment estimator. Asymptotic properties are given in both of these sections.
Section 5 presents a Monte Carlo study that compares the small sample performance of different methods.  Section 6 presents an application to a job training dataset with non-compliance. The paper concludes with a brief discussion in Section 7.

\section{Sample Complier Average Causal Effect}\label{sec:SCACE}

Consider a completely randomized experiment with $n$ units, where $n_1$ units are randomly assigned to treatment and the remaining $n_0$ units are assigned to control. Let $Z_i\in\{0,1\}$, $i=1,\cdots,n$, denote the indicator of treatment assigned for unit $i$.  Let $\bm{Z}$ denote the vector of $Z_i$ for the $n$ units.  Not everyone assigned treatment/control receives treatment/control, however.  Due to this, one can only estimate the average treatment effect under the restrictive assumption of homogeneous treatment effects. However, with heterogeneous effects it is possible to estimate the average treatment effect for those who comply with the treatment assignment.

We assume Stable Unit Treatment Value Assumption (SUTVA) \citep{Rubin_1978}, that is, there is only one version of treatment or control and there is no interference between units. Let $W_{i}(1)$ and $W_{i}(0)$, $i=1,\cdots,n$, be the potential treatment received if assigned to treatment and control for unit $i$.  Let $\bm{W}(z)$, $z=0,1$, be the vector of $W_i(z)$ for the $n$ units.  For $i=1,\cdots,n$, also define the four potential outcomes $Y_{i}(z,w)$ for which the treatment assigned and treatment received are fixed at $z=0,1$ and $w=0,1.$ For each individual only two of these potential outcomes, $Y_i(0)=Y_{i}(0,W_{i}(0))$ or $Y_i(1)=Y_{i}(1,W_{i}(1))$, can possibly be observed.  Let $\bm{Y}(z)$, $z=0,1$, be the vector of $Y_i(z)$ for the $n$ units. We further assume $Y_{i}(z,w)=Y_{i}(z^{\prime },w)$ for all $z,z^{\prime }$ and all $w$, that is, potential outcomes only depend on treatment received.  This is known as the exclusion restriction assumption \citep{Angrist_etal_1996}.

There are four possible latent types of units: always-takers (\textit{at}) who take treatment regardless of whether being assigned to treatment or control, with $W_i(1)=W_i(0)=1$; compliers (\textit{co}) take treatment when assigned to treatment and take control when assigned to control, with $W_i(1)=1$ and $W_i(0)=0$, or equivalently $W_i(1)-W_i(0)=1$; defiers (\textit{de}) who would take control when assigned to treatment and take treatment when assigned to control, with $W_i(1)=0$ and $W_i(0)=1$, or equivalently $W_i(1)-W_i(0)=-1$; and never-takers (\textit{nt}) who take control regardless of whether being assigned to treatment or control, with $W_i(1)=W_i(0)=0$.

If we regard units in the experiment as being independently and randomly sampled from a super-population, we can define the population complier average causal effect (CACE) as
\begin{equation*}
\tau_{CACE}^{pop}=\text{E}(Y_{i}(1)-Y_{i}(0)|W_{i}(1)-W_{i}(0)=1).
\label{pop_CACE}
\end{equation*}
The fraction of compliers in the super-population is $p_{co}^{pop}=\text{Pr}(W_i(1)-W_i(0)=1)$. \cite{Angrist_etal_1996} show that under further assumptions: (i) $P(W_i(1)=1)\neq P(W_i(0)=1)$ and (ii) $W_{i}(1)\geq W_{i}(0)$, it is possible to non-parametrically identify $\tau_{CACE}^{pop}$. Assumption (i) says that treatment assigned affects treatment received, and hence there exist compliers or defiers in the super-population. Assumption (ii) rules out defiers.  Under assumption (ii), $p_{co}^{pop}=\text{E}(W_i(1)-W_i(0))$.

In this paper we do not assume that units in the experiment are independently and randomly sampled from a super-population, and instead focus only on the finite population consisting of these units.  Thus, we treat $\bm{W}(0)$, $\bm{W}(1)$, $\bm{Y}(0)$ and $\bm{Y}(1)$ all as fixed, and treat only $\bm{Z}$ as random.

We rephrase assumptions (i) and (ii) in the finite population setting: (i') $W_i(1)\neq W_i(0)$ for some unit $i$ and (ii') $W_{i}(1)\geq W_{i}(0)$. Under these two assumptions, there is at least one complier and no defiers among the units in the study. Let $G_i,\ i=1,\cdots,n$ denote the latent group for unit $i$.  Since $W_i(1)$ and $W_i(0)$ are fixed, $G_i$ is also fixed.  The sample complier average causal effect (sample CACE) is defined as
\begin{equation*}
\tau _{CACE}=\frac{1}{n_{co}}\sum_{G_i=co}(Y_{i}(1)-Y_{i}(0)),
\label{samp_CACE}
\end{equation*}
where $n_{co}$ is the number of compliers in the sample. Under assumption (ii') $n_{co}=\sum_{i=1}^n (W_i(1)-W_i(0))$. The fraction of compliers in the sample is $p_{co}=n_{co}/n$. Further under the assumption that potential outcomes only depend on treatment received, for always-takers and never-takers $Y_i(1)-Y_i(0)=0$.  Therefore, $\sum_{G_i=co}(Y_{i}(1)-Y_{i}(0))=\sum_{i=1}^n (Y_{i}(1)-Y_{i}(0))$.

Define the sample average effect of treatment assigned on $W$ as
\begin{equation*}
\tau_{W}=\frac{1}{n}\sum_{i=1}^{n}(W_i(1)-W_i(0))=p_{co},
\label{samp_ITT_W}
\end{equation*}
and the sample average effect of treatment assigned on $Y$ as
\begin{equation*}
\tau_{Y}=\frac{1}{n}\sum_{i=1}^{n}(Y_{i}(1)-Y_{i}(0)).
\label{sample_ITT_Y}
\end{equation*}
We can then get another form of the sample CACE:
\begin{equation}
\tau _{CACE}=\frac{\tau_{Y}}{\tau_{W}}.
\label{samp_CACE2}
\end{equation}
$\tau _{CACE}$ is a parameter for the finite population of units in the experiment, for which point and interval estimates can be constructed.

\section{The Wald Estimator}

\subsection{Definition and Consistency of The Wald Estimator}
Let the observed indicator of treatment received be $W_{i}=W_{i}(Z_{i})$, $i=1,...,n$.  Let $\bm{W}$ be the vector of $W_i$ for all $n$ units.  Let the observed outcome be
$Y_{i}=Y_{i}(Z_{i})$. Let $\bm{Y}$ be the vector of $Y_i$ for all $n$ units.

For any set of quantities $Q_i$ for $i=1,\cdots,n$, let $\overline{Q}_1=\frac{1}{n_{1}}\sum_{i:Z_{i}=1}Q_{i}$ denote the mean for those assigned to treatment, let $\overline{Q}_{0}=\frac{1}{n_{0}}\sum_{i:Z_{i}=0}Q_{i}$ denote the mean for those assigned to control, and define the difference-in-means between these two groups as
\begin{equation*}
\widehat{\tau}_Q = \overline{Q}_1 -\overline{Q}_0.
\end{equation*}
Let $\widehat{\tau}_Y$ and $\widehat{\tau}_W$ denote the difference-in-means for $Y_i$ and $W_i$. When $\widehat{\tau}_W\neq 0$, the Wald estimator is defined as
\begin{equation}
\widehat{\tau}^{Wald}=\frac{\widehat{\tau}_{Y}}{\widehat{\tau}_{W}}=\frac{\overline{Y}_{1}-\overline{Y}_{0}}{%
\overline{W}_{1}-\overline{W}_{0}}.  \label{est_Wald}
\end{equation}

In the case when the units are regarded as being independently and randomly sampled from a superpopulation and treatment assignment is completely random, $\widehat{\tau}^{Wald}$ is a asymptotically unbiased and normally distributed estimator for $\tau_{CACE}^{pop}$, and the delta method can be used to estimate its asymptotic variance (see e.g. \cite{imbens2015causal}, Chapter 23). We will study the properties of using $\widehat{\tau}^{Wald}$ to estimate $\tau_{CACE}$.

The following theorem says that $\widehat{\tau}^{Wald}$ is a consistent estimator of $\tau_{CACE}$ under Condition S1 (details in the Supplemetary Material).
\begin{theorem}
	\label{consistency_CRE}
	Under Condition S1, $\widehat{\tau}^{Wald}-\tau_{CACE}=o_p(1)$ as $n\rightarrow\infty$.
\end{theorem}

\subsection{Construction of Confidence Interval Using the Wald Estimator}

For later use, we introduce some new notation.  For any set of quantities $\{Q_i:\ i=1,\cdots,n\}$, let $\overline{Q}=\frac{1}{n}\sum_{i=1}^{n}Q_i$ denote the mean for all $n$ units, and define the finite population variance as
\begin{equation*}
	\mathbb{S}_{Q}^2=\frac{1}{n-1}\sum_{i=1}^{n}\left(Q_i-\overline{Q}\right)^2.
\end{equation*}
For any two sets of quantities $\{Q_i:\ i=1,\cdots,n\}$ and $\{Q'_i:\ i=1,\cdots,n\}$, define the finite population covariance as
\begin{equation*}
	\mathbb{S}_{Q,Q'}=\frac{1}{n-1}\sum_{i=1}^{n}\left(Q_i-\overline{Q}\right)\left(Q_i'-\overline{Q}'\right).
\end{equation*}

To set our contribution into context we first discuss the approach given in \cite{li2017general}.

\subsubsection{The Approach in \cite{li2017general}}

\cite{li2017general} assumes that $Y_i(1)-Y_i(0)=\beta(W_i(1)-W_i(0))$ for all $i$. Under this linear instrumental variable model, for each complier, $W_i(1)-W_i(0)=1$ and hence $Y_i(1)-Y_i(0)=\beta$, which says that the individual treatment effect is homogeneous among compliers. It follows that $\tau_{CACE}=\beta$.

Define the adjusted outcome $A_i=Y_i-\beta W_i$ with potential outcomes $A_i(z)=Y_i(z)-\beta W_i(z)$ for $z=0,1$. Let $\widehat{\tau}_{A}$ be the difference-in-means for $A_i$.  Let $\mathbb{S}_{A}^2$, $\mathbb{S}_{A(1)}^2$ and $\mathbb{S}_{A(0)}^2$ be the finite population variances for $A_i$, $A_i(1)$ and $A_i(0)$. Let $\mathbb{S}_{A(1)-A(0)}^2$ be the finite population variance for $A_i(1)-A_i(0)$.  If $\{A_i:\ i=1,\cdots,n\}$ satisfy Condition (4) in Theorem 1 of \cite{li2017general}, then $\widehat{\tau}_A$ converges to a normal distribution with mean $\tau_{A}=\frac{1}{n}\sum_{i=1}^n(A_i(1)-A_i(0))$ and variance
\begin{equation*}
Var(\widehat{\tau}_A)=\frac{\mathbb{S}_{A(1)}^2}{n_1}+\frac{\mathbb{S}_{A(0)}^2}{n_0}-\frac{\mathbb{S}_{A(1)-A(0)}^2}{n}.
\end{equation*}

Because $Y_i(1)-\beta W_i(1)=Y_i(0)-\beta W_i(0)$ under the linear instrumental variable model, the adjusted outcome satisfies $A_i=A_i(1)=A_i(0)$. Therefore $\tau_{A}=0$, $\mathbb{S}_{A(1)}^2=\mathbb{S}_{A(0)}^2=\mathbb{S}_{A}^2$,  $\mathbb{S}_{A(1)-A(0)}=0$, and $Var(\widehat{\tau}_A)=(\frac{1}{n_1}+\frac{1}{n_0})\mathbb{S}_{A}^2$.
Let $\mathbb{S}_{Y}^2$ and $\mathbb{S}_{W}^2$ be the finite population variances for $Y_i$ and $W_i$, and let $\mathbb{S}_{Y,W}$ be the finite population covariance between $Y_i$ and $W_i$. Because $Y_i$ and $W_i$ are observed, $\mathbb{S}_{Y}^2$, $\mathbb{S}_{W}^2$ and $\mathbb{S}_{Y,W}$ are all observed. Due to the linear relationship between $A_i$ and $(Y_i,W_i)$, we have
\begin{equation*}
Var(\widehat{\tau}_A)=\left(\frac{1}{n_1}+\frac{1}{n_0}\right)\left(\mathbb{S}_{Y}^2+\beta^2\mathbb{S}_W^2-2\beta\mathbb{S}_{Y,W}\right).
\end{equation*}

Let $\nu_{1-\alpha/2}$ be the $(1-\alpha/2)$th quantile of the standard normal distribution.  Based on the normal approximation, the $1-\alpha$ confidence interval for $\beta$ consists of the values satisfying $|\widehat{\tau}_A/\sqrt{Var(\widehat{\tau}_A)}|\leq \nu_{1-\alpha/2}$, or $\widehat{\tau}_A^2\leq \nu_{1-\alpha/2}^2 Var(\widehat{\tau}_A)$. Due to the linear relationship between $A_i$ and $(Y_i,W_i)$, $\widehat{\tau}_A=\widehat{\tau}_Y-\beta\widehat{\tau}_W$. Therefore the $1-\alpha$ confidence interval for $\beta$ is the solution to the following inequality:
\begin{equation*}
\left(\widehat{\tau}_Y-\beta\widehat{\tau}_W\right)^2\leq \nu_{1-\alpha/2}^2\times\left(\frac{1}{n_1}+\frac{1}{n_0}\right)\left(\mathbb{S}_{Y}^2+\beta^2\mathbb{S}_W^2-2\beta\mathbb{S}_{Y,W}\right).
\end{equation*}
For different observed data, this solution can be an empty set, a finite interval $[c_1,c_2]$, the whole real line, or the union of two disjoint infinite intervals $(-\infty,c_1]\cup [c_2,\infty]$.  Infinite confidence sets suggest little or no identification due to weak instruments, and an empty set suggests possible violation of the linear instrumental variable model.

The linear instrumental variable model is overly restrictive. Also, the four possible forms of confidence sets are not handy for practical use. In the next sub-section we derive under heterogeneous effects a finite confidence interval for $\tau_{CACE}$  that are familiar to practitioners.

\subsubsection{A Finite Confidence Interval under Heterogeneous Effects}\label{sec:Wald:CI}

Define a new adjusted outcome $B_i=Y_i-\tau_{CACE} W_i$ with potential outcomes $B_i(z)=Y_i(z)-\tau_{CACE} W_i(z)$ for $z=0,1$.
Let $\widehat{\tau}_B$ be the difference-in-means for $B_i$. We have
\begin{equation}
\widehat{\tau}_B=\widehat{\tau}_Y-\tau_{CACE}\widehat{\tau_W}.\label{tauB_hat}
\end{equation}
Let $\mathbb{S}_{B(1)}^2$ and $\mathbb{S}_{B(0)}^2$ be the finite population variances for $B_i(1)$ and $B_i(0)$. Let $\mathbb{S}_{B(1)-B(0)}^2$ be the finite population variance for $B_i(1)-B_i(0)$.  Under Condition S1, we can apply Theorem 4 of \cite{li2017general} to obtain that $(\widehat{\tau}_B-\tau_B)/\sqrt{Var(\widehat{\tau}_B)}$ converges to the standard normal distribution, with $\tau_B=\frac{1}{n}\sum_{i=1}^n(B_i(1)-B_i(0))$ and
\begin{equation}
Var(\widehat{\tau}_B)=\frac{\mathbb{S}_{B(1)}^2}{n_1}+\frac{\mathbb{S}_{B(0)}^2}{n_0}-\frac{\mathbb{S}_{B(1)-B(0)}^2}{n}.\label{var_tauB}
\end{equation}
It is easy to see that $\tau_B = \tau_Y - \tau_{CACE}\tau_W$, which equals 0 according to \eqref{samp_CACE2}.  As $n\rightarrow\infty$, we have
\begin{equation*}
Pr\left(\widehat{\tau}_B - \nu_{1-\alpha/2}\sqrt{Var(\widehat{\tau}_B)}\leq 0 \leq \widehat{\tau}_B + \nu_{1-\alpha/2}\sqrt{Var(\widehat{\tau}_B)}\right)\rightarrow 1-\alpha.
\end{equation*}

Taking into account \eqref{tauB_hat} and \eqref{est_Wald}, when $\widehat{\tau}_W\neq 0$, we have
\begin{equation}
\label{ci_detail}
\begin{aligned}
&Pr\left(\widehat{\tau}_B - \nu_{1-\alpha/2}\sqrt{Var(\widehat{\tau}_B)}\leq 0 \leq \widehat{\tau}_B + \nu_{1-\alpha/2}\sqrt{Var(\widehat{\tau}_B)}\right)\\
=&Pr\left(\frac{\widehat{\tau }_{B}}{|\widehat{\tau}_W|} - \nu_{1-\alpha/2}\frac{\sqrt{Var(\widehat{\tau}_B)}}{|\widehat{\tau}_W|}\leq 0 \leq \frac{\widehat{\tau }_{B}}{|\widehat{\tau}_W|} + \nu_{1-\alpha/2}\frac{\sqrt{Var(\widehat{\tau}_B)}}{|\widehat{\tau}_W|}\right)\\
=&Pr\left(\frac{\widehat{\tau}_{Y}}{|\widehat{\tau}_W|}- \nu_{1-\alpha/2}\frac{\sqrt{Var(\widehat{\tau}_B)}}{|\widehat{\tau}_W|}\leq \frac{\widehat{\tau}_{W}}{|\widehat{\tau}_W|}\tau_{CACE}\leq\frac{\widehat{\tau}_{Y}}{|\widehat{\tau}_W|}+\nu_{1-\alpha/2}\frac{\sqrt{Var(\widehat{\tau}_B)}}{|\widehat{\tau}_W|}\right)\\
=&Pr\left(\widehat{\tau }^{Wald}-\nu_{1-\alpha/2}\frac{\sqrt{Var(\widehat{\tau}_B)}}{|\widehat{\tau}_W|}\leq\tau_{CACE}\leq\widehat{\tau }^{Wald}+\nu_{1-\alpha/2}\frac{\sqrt{Var(\widehat{\tau}_B)}}{|\widehat{\tau}_W|}\right),
\end{aligned}
\end{equation}
where the last equality can be obtained by considering each of the two cases with $\widehat{\tau}_W>0$ or $\widehat{\tau}_W<0$.
Therefore, as $n\rightarrow\infty$,
\begin{equation}
	\label{ci_CRE1}
	\widehat{\tau }^{Wald}\pm \nu_{1-\alpha/2}\sqrt{Var(\widehat{\tau}_B)}/|\widehat{\tau}_{W}|
\end{equation}
has $(1-\alpha)$ coverage rate for $\tau_{CACE}$.

If we can find an asymptotically conservative estimator for $Var\left(\widehat{\tau}_B\right)$, we can construct an asymptotically conservative $(1-\alpha)$ confidence interval for $\tau_{CACE}$.  Replacing $\tau_{CACE}$ in the expression of $B_i$ with its estimator $\widehat{\tau}^{Wald}$, we define $\widehat{B}_i=Y_{i}-\widehat{\tau}^{Wald}W_i$, which is observed.
We can use the sample variance of $\widehat{B}_i$ for those assigned to treatment arm $z$,
\begin{equation*}
S_{\widehat{B}(z)}^2=\frac{1}{n_z-1}\sum_{i:Z_i=z}\left(\widehat{B}_i-\sum_{i:Z_i=z}\widehat{B}_i/n_z\right)^2,
\end{equation*}
to estimate $\mathbb{S}_{B(z)}^2$ for $z=0,1$.  We propose the following estimator for $Var\left(\widehat{\tau}_B\right)$:
\begin{equation}		
\widehat{Var}(\widehat{\tau}_B)=\frac{S_{\widehat{B}(1)}^2}{n_1}+\frac{S_{\widehat{B}(0)}^2}{n_0}.\label{Var_tauB_hat}
\end{equation}

The following theorem says that the confidence interval constructed using $\widehat{Var}(\widehat{\tau}_B)$ is asymptotically conservative under Condition S1.
\begin{theorem}
\label{conservative_CRE}
Under Condition S1, the confidence interval
\begin{equation}
	\label{ci_CRE2}
	\widehat{\tau}^{Wald}\pm \nu_{1-\alpha/2}\sqrt{\widehat{Var}(\widehat{\tau}_B)}/\left|\widehat{\tau}_{W}\right|
\end{equation}
has at least $(1-\alpha)$ coverage rate for $\tau_{CACE}$ as $n\rightarrow\infty$.
\end{theorem}

Proposition S1 in the Supplemetary Material shows that this confidence interval is the same as the super-population confidence interval for $\tau_{CACE}^{pop}$ obtained by the delta method.

\section{The Regression Adjustment Estimator}
\subsection{Definition and Consistency of The Regression Adjustment Estimator}
The Wald estimator does not account for any covariates.  In practical experiments, covariates are often available, and we can adjust for them to improve efficiency.

Let $\bm{x}_i$, $i=1,\cdots,n$, be the $K\times 1$ vector of covariates for unit $i$. These covariates are all observed before treatment assignment and are treated as fixed. Without loss of generality, we assume that the covariates have been centered around zero, that is, $\overline{\bm{x}}\equiv 1/n\sum_{i=1}^n \bm{x}_i=\bm{0}$. Let $\bm{X}$ be the $n\times K$ matrix of covariates for the $n$ units.

For any set of quantities $\{Q_i:\ i=1,\cdots,n\}$, let $\widehat{\alpha}_{Q,z}$ and $\widehat{\bm{\beta}}_{Q,z}$ be the intercept and the vector of coefficients on $\bm{x}_i$ in the linear projection of $Q_i$ on $\bm{x}_i$ using units with $Z_i=z$ for $z=0,1$. That is,
\begin{equation*}
	\left(\widehat{\alpha}_{Q,z},\widehat{\bm{\beta}}_{Q,z}\right)=\arg\min_{\alpha,\bm{\beta}}\sum_{i:\ Z_i=z}\left(Q_i-\alpha-\bm{\beta}^{\top}\bm{x}_i\right)^2.
\end{equation*}
For quantities $\{W_i:\ i=1,\cdots,n\}$, $\{Y_i:\ i=1,\cdots,n\}$ and $\{B_i:\ i=1,\cdots,n\}$, we can thus define $\widehat{\bm{\beta}}_{W,z}$, $\widehat{\bm{\beta}}_{Y,z}$ and $\widehat{\bm{\beta}}_{B,z}$ ($z=0,1)$. Note that, $\widehat{\bm{\beta}}_{W,z}$ and $\widehat{\bm{\beta}}_{Y,z}$ are obtained by the Ordinary Least squares (OLS) regression of $Y_i$ and $W_i$ on $\bm{x}_i$ using units with $Z_i=z$.

According to the results of \cite{lin2013agnostic}, under complete randomization, $\tau_W$ or $\tau_Y$ can be consistently estimated by the estimated coefficient on $Z_i$ (denoted by $\widehat{\tau}_W^{Reg}$ or $\widehat{\tau}_Y^{Reg}$) in the OLS regression of $W_i$ or $Y_i$ on $Z_i$, $\bm{x}_i$ and $Z_i\bm{x}_i$ using all $n$ units.

Define $\overline{\bm{x}}_1=\sum_{i:\ Z_i=1}\bm{x}_i/n_1$ and $\overline{\bm{x}}_0=\sum_{i:\ Z_i=0}\bm{x}_i/n_0$. Applying Proposition S2 in the Supplemetary Material, we can write $\widehat{\tau}_W^{Reg}$ and $\widehat{\tau}_Y^{Reg}$ as
\begin{equation}
\begin{aligned}
\widehat{\tau}_W^{Reg}&=\frac{1}{n_1}\sum_{i:\ Z_i=1}\left(W_i-\widehat{\bm{\beta}}_{W,1}\bm{x}_i\right)-\frac{1}{n_0}\sum_{i:\ Z_i=0}\left(W_i-\widehat{\bm{\beta}}_{W,0}\bm{x}_i\right)\\
&=\widehat{\tau}_W - \widehat{\bm{\beta}}_{W,1}\overline{\bm{x}}_1+\widehat{\bm{\beta}}_{W,0}\overline{\bm{x}}_0
\end{aligned}\label{ITThat_W_adj}
\end{equation}
and
\begin{equation}
\begin{aligned}
\widehat{\tau}_Y^{Reg}&=\frac{1}{n_1}\sum_{i:\ Z_i=1}\left(Y_i-\widehat{\bm{\beta}}_{Y,1}\bm{x}_i\right)-\frac{1}{n_0}\sum_{i:\ Z_i=0}\left(Y_i-\widehat{\bm{\beta}}_{Y,0}\bm{x}_i\right)\\
&=\widehat{\tau}_Y - \widehat{\bm{\beta}}_{Y,1}\overline{\bm{x}}_1+\widehat{\bm{\beta}}_{Y,0}\overline{\bm{x}}_0.
\end{aligned}\label{ITThat_Y_adj}
\end{equation}
The following theorem says that the estimator,
\begin{equation}
\widehat{\tau}^{Reg}=\frac{\widehat{\tau}_{Y}^{Reg}}{\widehat{\tau}_{W}^{Reg}},  \label{est_adj}
\end{equation}
is a consistent estimator of $\tau_{CACE}$ under condition Condition S2 (details in the Supplemetary Material).
\begin{theorem}
\label{consistency_CRE_adj}
	Under Condition S2, $\widehat{\tau}^{Reg}-\tau_{CACE}=o_p(1)$ as $n\rightarrow\infty$.
\end{theorem}

\subsection{Construction of Confidence Interval Using the Regression Adjustment Estimator}
\label{sec:reg:ci}

Because $B_i$ is a linear combination of $Y_i$ and $W_i$, we have
\begin{equation}\label{beta_hat_Bz}
	\widehat{\bm{\beta}}_{B,z}=\widehat{\bm{\beta}}_{Y,z}-\tau_{CACE}\widehat{\bm{\beta}}_{W,z}.
\end{equation}
Define a new adjusted outcome
\begin{equation*}
D_i=B_i-\widehat{\bm{\beta}}_{B,Z_i}^{\top}\bm{x}_i,
\end{equation*}
with potential outcomes $D_i(z)=B_i(z)-\widehat{\bm{\beta}}_{B,z}^{\top}\bm{x}_i$. Let $\widehat{\tau}_D$ be the difference-in-means for $D_i$. Let $Var(\widehat{\tau}_D)$ denote the variance of $\widehat{\tau}_D$. The following theorem says that $\widehat{\tau}_D$ asymptotically has a normal distribution under Condition S2.
\begin{theorem}
	\label{asymnormal_CRE_adj}
	Under Condition S2, as $n\rightarrow\infty$,
	\begin{equation}
	\widehat{\tau}_D/\sqrt{Var(\widehat{\tau}_D)}\stackrel{d}{\longrightarrow}N(0,1).
	\end{equation}
\end{theorem}

With Theorem~\ref{asymnormal_CRE_adj}, as $n\rightarrow\infty$, we have
\begin{equation*}
Pr\left(\widehat{\tau}_D - \nu_{1-\alpha/2}\sqrt{Var(\widehat{\tau}_D)}\leq 0 \leq \widehat{\tau}_D + \nu_{1-\alpha/2}\sqrt{Var(\widehat{\tau}_D)}\right)\rightarrow 1-\alpha.
\end{equation*}
Combining the definitions of $D_i$ and $B_i$ with \eqref{ITThat_W_adj}, \eqref{ITThat_Y_adj} and \eqref{beta_hat_Bz}, we have
\begin{equation}\label{tauD_hat}
\begin{aligned}
&\widehat{\tau}_D=\frac{1}{n_1}\sum_{i:\ Z_i=1}D_i-\frac{1}{n_0}\sum_{i:\ Z_i=0}D_i\\
=&\frac{1}{n_1}\sum_{i:\ Z_i=1}\left(B_i-\widehat{\bm{\beta}}_{B,1}^{\top}\bm{x}_i\right)-
\frac{1}{n_0}\sum_{i:\ Z_i=0}\left(B_i-\widehat{\bm{\beta}}_{B,0}^{\top}\bm{x}_i\right)\\
=&\frac{1}{n_1}\sum_{i:\ Z_i=1}\left(Y_i-\tau_{CACE}W_i-\left(\widehat{\bm{\beta}}_{Y,1}-\tau_{CACE}\widehat{\bm{\beta}}_{W,1}\right)^{\top}\bm{x}_i\right)\\
&-
\frac{1}{n_0}\sum_{i:\ Z_i=0}\left(Y_i-\tau_{CACE}W_i-\left(\widehat{\bm{\beta}}_{Y,0}-\tau_{CACE}\widehat{\bm{\beta}}_{W,0}\right)^{\top}\bm{x}_i\right)\\
=&\widehat{\tau}_Y^{Reg}-\tau_{CACE}\widehat{\tau}_W^{Reg}.
\end{aligned}
\end{equation}
Note that $\widehat{\tau}_D$ has a similar form as $\widehat{\tau}_B$ in \eqref{tauB_hat}.  Following similar arguments to those in Section~\ref{sec:Wald:CI} that lead to \eqref{ci_CRE1} having $(1-\alpha)$ coverage rate for $\tau_{CACE}$, we obtain that as $n\rightarrow\infty$,
\begin{equation*}
	\label{ci_Reg}
	\widehat{\tau}^{Reg}\pm \nu_{1-\alpha/2}\sqrt{Var(\widehat{\tau}_D)}/\left|\widehat{\tau}_{W}^{Reg}\right|
\end{equation*}
has $(1-\alpha)$ coverage rate for $\tau_{CACE}$.

If we can find an asymptotically conservative estimator for $Var\left(\widehat{\tau}_D\right)$, we can construct an asymptotically conservative $(1-\alpha)$ confidence interval for $\tau_{CACE}$.  Applying Proposition S2 in the Supplementary Material, we know that $\widehat{\tau}_D$ equals the coefficient on $Z_i$ in the linear projection of $B_i$ on $Z_i$, $\bm{x}_i$ and $Z_i\bm{x}_i$ using all $n$ units.  If $B_i$ were known, we could use the EHW robust variance estimator to estimate $Var(\widehat{\tau}_D)$.  However, as $B_i$ contains the unknown quantity $\tau_{CACE}$, in practice we replace $\tau_{CACE}$ with $\widehat{\tau}^{Reg}$, and consider instead the linear projection of
\begin{equation*}
	\widehat{B}_i=Y_i-\widehat{\tau}^{Reg}W_i,
\end{equation*}
that is, the OLS regression of $\widehat{B}_i$ since $\widehat{B}_i$ is observed.

Let $\widehat{Var}^{EHW}(\widehat{\tau}_D)$ denote the EHW variance estimator of the estimated coefficient on $Z_i$ in the OLS regression of $\widehat{B}_i$ on $Z_i$, $\bm{x}_i$ and $Z_i\bm{x}_i$ using all $n$ units. We can also use the series of HC variance estimators that can be used as finite sample corrections of the EHW variance estimator in usual OLS regressions \citep{mackinnon2012thirty}. Specifically, let $\widehat{Var}^{HC2}(\widehat{\tau}_D)$ and $\widehat{Var}^{HC3}(\widehat{\tau}_D)$ denote the HC2 and HC3 variance estimators of the estimated coefficient on $Z_i$ in the OLS regression of $\widehat{B}_i$ on $Z_i$, $\bm{x}_i$ and $Z_i\bm{x}_i$ using all $n$ units.  The following theorem says that the confidence interval constructed using $\widehat{Var}^{EHW}(\widehat{\tau}_D)$, $\widehat{Var}^{HC2}(\widehat{\tau}_D)$ or $\widehat{Var}^{HC3}(\widehat{\tau}_D)$ is asymptotically conservative under Condition S2.
\begin{theorem}
\label{conservative_reg}
Under Condition S2, each of the following confidence intervals,
\begin{align}
	&\widehat{\tau}^{Reg}\pm \nu_{1-\alpha/2}\sqrt{\widehat{Var}^{EHW}(\widehat{\tau}_D)}/\left|\widehat{\tau}_{W}^{Reg}\right|,		\label{ci_reg_EHW}\\
	&\widehat{\tau}^{Reg}\pm \nu_{1-\alpha/2}\sqrt{\widehat{Var}^{HC2}(\widehat{\tau}_D)}/\left|\widehat{\tau}_{W}^{Reg}\right|,		\label{ci_reg_HC2}\\
\text{and }	&\widehat{\tau}^{Reg}\pm \nu_{1-\alpha/2}\sqrt{\widehat{Var}^{HC3}(\widehat{\tau}_D)}/\left|\widehat{\tau}_{W}^{Reg}\right|,		\label{ci_reg_HC3}
\end{align}
is asymptotically equivalent and has at least $(1-\alpha)$ coverage rate for $\tau_{CACE}$ as $n\rightarrow\infty$.
\end{theorem}

\subsection{Efficiency Gain Using Regression Adjustment}
To characterize the efficiency gain using regression adjustment, we investigate the percentage by which regression adjustment reduces the length of the asymptotic $1-\alpha$ confidence interval, which we refer to as the percent reduction in length (PRIL).

For $z=0,1$, let $\mathbb{S}_{B(z)|\bm{x}}^2$ denote the finite population variance of the linear projection of $B_i(z)$ on $\bm{x}_i$, $\alpha_{B,z}+\bm{\beta}_{B,z}^{\top}\bm{x}_i$, where $\alpha_{B,z}$ and $\bm{\beta}_{B,z}$ are the intercept and the vector of coefficients on $\bm{x}_i$ in the linear projection of $B_i(z)$ on $\bm{x}_i$ using all $n$ units.  Define
\begin{align} Var(\widehat{\tau}_B)^{+}&=\frac{\mathbb{S}_{B(1)}^2}{n_1}+\frac{\mathbb{S}_{B(0)}^2}{n_0},\label{variance_estimand_nox}\\ Var(\widehat{\tau}_{B\mid\bm{x}})^{+}&=\frac{\mathbb{S}_{B(1)\mid\bm{x}}^2}{n_1}+\frac{\mathbb{S}_{B(0)\mid\bm{x}}^2}{n_0},\label{variance_estimand}
\end{align}
\begin{theorem}
	Under Condition S2, compared with \eqref{ci_CRE2}, the PRIL of \eqref{ci_reg_EHW}, \eqref{ci_reg_HC2} or \eqref{ci_reg_HC3} converges in probability to
	\begin{equation*}
		1-\sqrt{1-Var(\widehat{\tau}_{B\mid\bm{x}})^{+}/Var\left(\widehat{\tau}_B\right)^{+}}.
	\end{equation*}
\end{theorem}

For $z=0,1$, since $\mathbb{S}_{B(z)|\bm{x}}^2$ is the finite population variance of the linear projection of $B_i(z)$ on $\bm{x}_i$, we have $\mathbb{S}_{B(z)\mid\bm{x}}^2\leq{}\mathbb{S}_{B(z)}^2$. Hence $Var(\widehat{\tau}_{B\mid\bm{x}})^{+}\leq
Var(\widehat{\tau}_B)^{+}$, and the PRIL is asymptotically nonnegative. Furthermore, recall that $B_i(z)=Y_i(z)-\tau_{CACE}W_i(z)$. If the covariates are more predictive of $Y_i(z)$ or $W_i(z)$, $\mathbb{S}_{B(z)\mid\bm{x}}^2$ is closer to $\mathbb{S}_{B(z)}^2$, then $Var(\widehat{\tau}_{B\mid\bm{x}})^{+}$ is closer to $Var(\widehat{\tau}_B)^{+}$, and the limit of the PRIL is larger.

\section{Monte Carlo Simulation}
The focus of the Monte Carlo simulation is to compare the performance of the methods.  We use ``Wald" to denote the Wald estimator. For confidence intervals constructed using the Wald estimator, we use ``Wald-LD" to denote the confidence set constructed using the approach in \cite{li2017general}, and use ``Wald-Delta" to denote the confidence interval constructed using our approach.  We use ``Reg" to denote the regression adjustment estimator. For confidence intervals constructed using the regression adjustment estimator, we use ``Reg-EHW", ``Reg-HC2" and ``Reg-HC3" to denote the confidence intervals constructed using the EHW, HC2 or HC3 variance estimators.
	
\subsection{Setup of the Monte Carlo Simulation}\label{sec:setup_MCstudy}
We set $n_1=n_0=n/2$. Each covariate independently follows a $N(0, 1)$ distribution. Noting that potential outcomes only depend on treatment received, we use $Y_i^W(w)$ ($w=0,1$) to denote the potential outcome if receiving treatment arm $w$. The potential outcomes and the potential latent variables are generated using the following data generating process:
\begin{equation*}
\begin{aligned}
	Y_i^W(0) &=\bm{\xi}^{\top}\bm{x}_i+\epsilon_{0 i}, \\
	Y_i^W(1) &=\bm{\xi}^{\top}\bm{x}_i+\bm{\eta}^{\top}\bm{x}_i+\epsilon_{1 i},\\
	L_i(0) &=\delta_0+\bm{\psi}^{\top}\bm{x}_i+u_i, \\
	L_i(1) &=\delta_1+L_i(0).
\end{aligned}
\end{equation*}
We then set $W_i(z)=I(L_i(z)>0)$ for $z=0,1$. Here $\bm{\xi}$, $\bm{\eta}$ and $\bm{\psi}$ are each a $K\times 1$ vector with all elements being 1.

For a given value of $p_{co}$, we set $\delta_0=((p_{co}-0.5)/0.35+1)\sqrt{K}$, and choose the value of $\delta_1$ such that the fraction of compliers equals $p_{co}$.  We fix $K=5$, and consider $n\in \{200,400\}$ and $p_{co}\in \{0.85,0.5,0.15\}$.

The error terms are generated according to
	$$
	\left(\begin{array}{c}
		\kappa_0\epsilon_{i 0} \\
		\kappa_1\epsilon_{i 1} \\
		\kappa_2u_{i}
	\end{array}\right) \sim N\left(\bm{0},\left(\begin{array}{ccc}
		1 & 0 & \rho \\
		0 & 1 & \rho \\
		\rho & \rho & 1
	\end{array}\right)\right)
	$$
with $\rho\in \{0,0.5\}$.  The values of $\kappa_0$, $\kappa_1$ and $\kappa_2$ are chosen such that the squared multiple correlation coefficient is $0.5$ in each of the equations for $Y_i^W(0)$, $Y_i^W(1)$ and $L_i(0)$.

The values of $(\bm{x}_i, L_i(0), L_i(1), W_i(0), W_i(1), Y_i(0), Y_i(1))$ are fixed, and only the values of $Z_i$ are randomly generated $1,000$ times under complete randomization. The observed values are $(\bm{x}_i, Z_i, W_i, Y_i)$.  We thus have $1,000$ datasets for each of 12 scenarios with different combinations of $n$, $p_{co}$ and error distribution.

\subsection{Performance Measures}

If $\widehat{\tau}_W=0$ or $\widehat{\tau}_W^{Reg}=0$, the point estimate is infinite, and the confidence interval is also infinite, regardless of which method is used to constructed the confidence interval.  These results are regarded as being abnormal. In addition, for the Wald-LD method, we view the three forms of confidence sets other than finite confidence intervals as being abnormal. We remove these results in the following comparison.

For the $r$th dataset and the $m$th method, let $\widehat{\tau}_{r m}$ denote the point estimate, let $\mathcal{C}_{r m}$ denote the $95\%$ interval estimate, and let $L_{r m}$ denote the length of $\mathcal{C}_{r m}$.  We consider three performance measures for method $m$: (1) median absolute error
\begin{equation*}
		\text{MAE}_{m} =\text{median of }|\widehat{\tau}_{r m}-\tau_{CACE}|,
\end{equation*}
(2) coverage rate of 95\% intervals
\begin{equation*}
		\text{CRate}_{m} =\text{mean of }I\left(\tau_{CACE} \in \mathcal{C}_{r m}\right),
\end{equation*}
and (3) median length of 95\% intervals
\begin{equation*}
\text{Len}_m = \text{median of }L_{rm}.
\end{equation*}
	
\subsection{Results}

In our simulation, abnormal results only appear when $p_{co}=0.15$.  Table~\ref{T:DEL} presents the proportions of abnormal confidence intervals for different methods when $p_{co}=0.15$.  The Wald-LD method has rather high proportions of abnormal confidence intervals, all in the form of the whole real line or the union of two disjoint infinite intervals, indicating poor identification due to weak instruments. These proportions are especially high when the sample size is small.  The Wald-Delta method has zero or very low proportion of abnormal results with $\widehat{\tau}_W=0$.  The regression adjustment methods have no abnormal results with $\widehat{\tau}_W^{Reg}=0$.

\begin{figure}
  \centering
  \subfigure[point estimates]{\includegraphics[width=10cm]{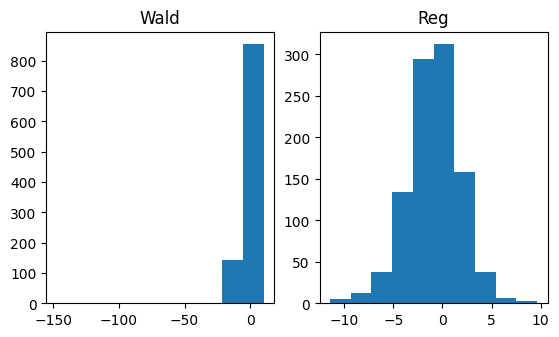}}
  \subfigure[lengths of 95\% intervals]{\includegraphics[width=10cm]{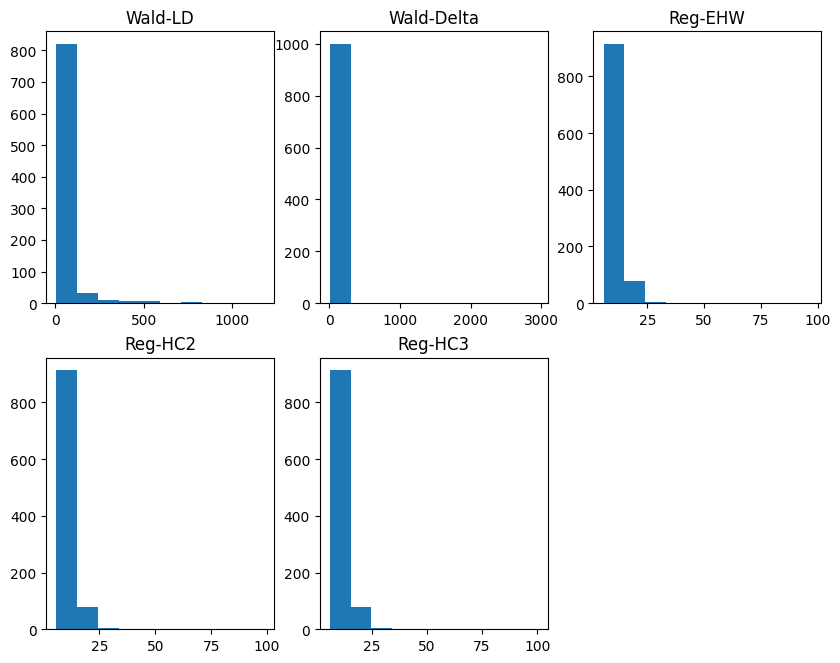}}
  \caption{Histograms of point estimates and lengths of 95\% intervals with $n=400$, $p_{co}=0.15$ and $\rho=0$.}
  \label{fig:hist}
\end{figure}

When $\widehat{\tau}_W$ or $\widehat{\tau}_W^{Reg}$ is nonzero but close to zero, the Wald method or the regression adjustment method can be unstable.  The point estimate can be quite large in absolute value, and the corresponding confidence interval can be quite wide.  To illustrate this point, for the 1,000 data sets generated with $n=400$, $p_{co}=0.15$ and $\rho=0$, Figure~\ref{fig:hist} presents histograms of point estimates and lengths of 95\% confidence intervals, with abnormal results excluded.  It appears that the regression adjustment method is much stabler than the Wald method.  The histogram of lengths of the 95\% confidence intervals has a longer tail for the Wald-Delta method than for the Wald-LD method.  This is because for the Wald-LD method 13.1\% of the datasets with infinite intervals have been excluded, whereas for the Wald-Delta method no datasets with infinite intervals have been excluded.

With abnormal results excluded, Table~\ref{T:MAE} presents median absolute error, Table~\ref{T:CR} presents coverage rate of 95\% intervals measured in percentage differences from the nominal level 0.95, $100\times (\text{CRate}_m-0.95)$, and   Table~\ref{T:LEN} presents median length of 95\% intervals.

	\begin{table}[tbp]
			\caption{Proportions of abnormal confidence intervals for different methods when $p_{co}=0.15$.}
			\label{T:DEL}
			\begin{center}
				\begin{tabular}{lcc}
					method & {$\rho=0$} & {$\rho=0.5$} \\
					\hline
					& \multicolumn{2}{c}{$n=200$}\\
					Wald-LD  &  0.395 &  0.425 \\
					Wald-Delta &  0.004 &  0.004\\
					Reg-EHW, Reg-HC2, Reg-HC3 &  0.000 &  0.000\\
					\hline
					& \multicolumn{2}{c}{$n=400$}\\
					Wald-LD   &  0.131 &  0.119\\
					Wald-Delta &  0.000 &  0.000\\
					Reg-EHW, Reg-HC2, Reg-HC3   &  0.000 &  0.000\\
				\end{tabular}
			\end{center}
	\end{table}

	\begin{table}[tbp]
			\caption{Median absolute error for different methods.}
			\label{T:MAE}
			\begin{center}
				\begin{tabular}{lcccccc}
					& \multicolumn{2}{c}{$p_{co}=0.85$} & \multicolumn{2}{c}{$p_{co}=0.5$} & \multicolumn{2}{c}{$p_{co}=0.15$}\\
					method & {$\rho=0$} & {$\rho=0.5$} & {$\rho=0$} & {$\rho=0.5$} & {$\rho=0$} & {$\rho=0.5$}\\
					\hline
					& \multicolumn{6}{c}{$n=200$}\\
					Wald      &  0.450 &  0.517 &  0.855 &  0.897 &  3.031 &  3.028\\
					Reg &  0.301 &  0.362 &  0.591 &  0.643 &  2.205 &  2.239 \\
					\hline
					& \multicolumn{6}{c}{$n=400$}\\
					Wald      &  0.377 &  0.350 &  0.778 &  0.682 &  2.597 &  2.288\\
					Reg &  0.245 &  0.222 &  0.484 &  0.419 &  1.749 &  1.557\\
				\end{tabular}
			\end{center}
	\end{table}

	\begin{table}[tbp]
			\caption{Coverage rate of 95\% intervals for different methods, measured in percentage differences from the nominal level 0.95, $100\times (\text{CRate}_m-0.95)$.}
			\label{T:CR}
			\begin{center}
				\begin{tabular}{lcccccc}
					& \multicolumn{2}{c}{$p_{co}=0.85$} & \multicolumn{2}{c}{$p_{co}=0.5$} & \multicolumn{2}{c}{$p_{co}=0.15$}\\
					method & {$\rho=0$} & {$\rho=0.5$} & {$\rho=0$} & {$\rho=0.5$} & {$\rho=0$} & {$\rho=0.5$}\\
					\hline
					& \multicolumn{6}{c}{$n=200$}\\
					Wald-LD       &  5.0 &  5.0 &  5.0 &  5.0 &  5.0 &  5.0\\
					Wald-Delta      &  2.4 &  1.4 &  1.3 &  1.4 &  -0.2 &  0.3\\
					Reg-EHW &  4.1 &  3.0 &  1.8 &  2.5 &  3.3 &  3.6\\
					Reg-HC2 &  4.4 &  3.5 &  2.1 &  2.7 &  3.5 &  4.0\\
					Reg-HC3 &  4.5 &  3.7 &  2.2 &  3.1 &  3.7 &  4.1\\
					\hline
					& \multicolumn{6}{c}{$n=400$}\\
					Wald-LD       &  5.0 &  5.0 &  5.0 &  5.0 &  5.0 &  5.0\\
					Wald-Delta      &  2.5 &  3.3 &  0.6 &  2.5 &  2.5 &  1.5\\
					Reg-EHW &  3.6 &  3.4 &  2.0 &  2.8 &  2.9 &  2.7\\
					Reg-HC2 &  3.9 &  3.7 &  2.3 &  3.0 &  3.4 &  2.9\\
					Reg-HC3 &  4.1 &  3.8 &  2.3 &  3.2 &  3.5 &  3.0\\
				\end{tabular}
			\end{center}
	\end{table}

	\begin{table}[tbp]
			\caption{Median length of 95\% intervals for different methods.}
			\label{T:LEN}
			\begin{center}
				\begin{tabular}{lcccccc}
					& \multicolumn{2}{c}{$p_{co}=0.85$} & \multicolumn{2}{c}{$p_{co}=0.5$} & \multicolumn{2}{c}{$p_{co}=0.15$}\\
					method & {$\rho=0$} & {$\rho=0.5$} & {$\rho=0$} & {$\rho=0.5$} & {$\rho=0$} & {$\rho=0.5$}\\
					\hline
					& \multicolumn{6}{c}{$n=200$}\\
					Wald-LD       &  3.486 &  4.027 &  6.477 &  7.124 &  30.035 &  30.327\\
					Wald-Delta      &  3.079 &  3.504 &  5.341 &  5.906 &  18.367 &  18.722\\
					Reg-EHW &  2.231 &  2.403 &  3.846 &  4.021 &  13.091 &  13.722\\
					Reg-HC2 &  2.298 &  2.490 &  3.966 &  4.163 &  13.494 &  14.233\\
					Reg-HC3 &  2.369 &  2.580 &  4.090 &  4.311 &  13.906 &  14.790\\
					\hline
					& \multicolumn{6}{c}{$n=400$}\\
					Wald-LD       &  2.828 &  2.721 &  5.339 &  4.965 &  23.131 &  21.812 \\
					Wald-Delta      &  2.434 &  2.404 & 4.421 &  4.195 &  14.165 &  13.573\\
					Reg-EHW &  1.711 &  1.676 &  3.103 &  2.925 &   10.697 &   9.532\\
					Reg-HC2 &  1.738 &  1.702 &  3.151 &  2.971 &   10.864 &   9.675\\
					Reg-HC3 &  1.765 &  1.730 &  3.201 &  3.018 &   11.033 &   9.820\\
				\end{tabular}
			\end{center}
	\end{table}

In term of the median absolute error, compared to the Wald method, regression adjustment has a reduction of 26.1\% to 38.6\% across different scenarios. In terms of coverage rate, none of the methods has under coverage. However, the Wald-LD method is too conservative as its coverage rates are 100\% in all scenarios. In terms of median length of 95\% intervals, (i) compared to the Wald-LD method, the Wald-Delta method has a reduction of 11.7\% to 38.9\%; (ii) compared to the Wald-Delta method, the Reg-EHW method has a reduction of 24.5\% to 31.9\%; (iii) compared to the Reg-EHW method, the Reg-HC2 method has an increase of 1.5\% to 3.7\%, and the Reg-HC3 method has an increase of 3.0\% to 7.8\%.

To summarize, we recommend the Wald-Delta method when covariates are unavailable, and the Reg-EHW method when covariates are available.

\section{Application to a Job Training Experiment with Non-compliance}
\label{application}
The Job Search Intervention Study (JOBS II) dataset comes from a field experiment designed and conducted by \cite{vinokur1995impact} that investigates the efficacy of a job training intervention on unemployed workers. Participants were randomly selected to attend the JOBS II training program that taught job-search skills and coping strategies for dealing with setbacks in the job-search process.  In the dataset, 372 participants were selected to attend the training program and actually attended, 228 participants were selected to attend the training program but did not attend, and 299 participants were not selected to attend the training program and did not attend. One outcome of interest is a continuous variable measuring the level of job-search self-efficacy with values from 1 to 5.  The covariates include age, gender, ethnicity, marital status, monthly income and educational attainment.  This data set has been analyzed in Chapter 21 of \cite{ding2023first}.

Since no one could attend the training program without being selected, there were no always-takers or defiers. Those who were selected to attend the training program and actually attended were compliers, those who were selected to attend the training program but did not attend were never-takers, and those who were not selected to attend the training program and did not attend were a mixture of compliers and never-takers. Obviously assumptions (i') and (i'') in Section~\ref{sec:SCACE} are satisfied.

For different methods presented in this paper, Table~\ref{real_data} presents point estimates of $p_{co}$, and point estimates and 95\% intervals of $\tau_{CACE}$.  The estimated fraction of compliers is around 0.62 for all methods.  The points estimates of $\tau_{CACE}$ from different methods are similar.  The Reg-EHW method has a slightly shorter 95\% interval than the Wald-Delta method.  The Reg-HC2 and Reg-HC3 methods have slightly longer 95\% intervals than the Wald-Delta method. The Wald-LD method has a much longer 95\% interval than other methods, which is overly conservative.
	
	\begin{table}[htbp]
		\centering
		\caption{Point estimates of $p_{co}$, and point estimates and 95\% intervals of $\tau_{CACE}$ from different methods for the JOBS II dataset}
		\label{real_data}
		\begin{tabular}{lccc}
            & point estimate & point estimate & 95\% interval\\
			method  & of $p_{co}$  & of $\tau_{CACE}$ &    of $\tau_{CACE}$\\
			\hline
			Wald-LD & 0.620 & 0.109 &  [-0.141, 0.358] \\
			Wald-Delta  & 0.620 & 0.109 &  [-0.050, 0.268] \\
			Reg-EHW &  0.616 & 0.118 &  [-0.039, 0.274] \\
			Reg-HC2 & 0.616 & 0.118 &  [-0.042, 0.278] \\
			Reg-HC3 & 0.616 & 0.118 &  [-0.046, 0.281] \\
		\end{tabular}
	\end{table}
	
\section{Discussion}
In pragmatic randomized experiments, incomplete adherence and/or incomplete compliance to the assigned treatment is common. This paper focuses on inference about the sample complier average causal effect.  We propose to use the Wald estimator and the regression adjustment estimator, and also propose methods for construction of confidence intervals using these two estimators.  We establish their asymptotic properties, and compare their small sample performance using a Monte Carlo simulation.

The results from the Monte Carlo simulation show that when covariates are available, compared to the Wald method, the regression adjustment method is more stable, does not yield abnormal results, and reduces median absolute error of point estimates and median length of 95\% confidence intervals, without sacrificing coverage rate of the confidence intervals.  When covariates are not available, in terms of construction of confidence intervals based on the Wald estimator, our proposed method yields a lot less abnormal results and greatly reduces median length of 95\% confidence intervals, compared to a previous method in \cite{li2017general}.  We also present an application to the Job Search Intervention Study (JOBS II) dataset.

\bigskip
\begin{center}
{\large\bf SUPPLEMENTARY MATERIAL}
\end{center}

\begin{description}

\item[Conditions and Proofs for Theorems and Propositions:] Including Conditions S1 and S2, Propositions S1 and S2 and their proof, and proof for Theorem 1-6. (.pdf file)

\item[R-package SCACE:] The R package SCACE contains code that implements the Wald estimator (``WaldLD.R" and ``WaldDelta.R") and the regression adjustment estimator (``Reg.R"). It also contains code for reproducing the Monte Carlo simulation (``genData.R" and ``mc.R") and code for analyzing the JOBS II dataset (``example.R", which uses the ``jobs" dataset in R package ``mediation"). A readme document is also included. (.zip file)

\end{description}

\section*{Supplementary Material: Conditions for Theorems and Propositions}\label{sec:conditions}
\setcounter{theorem}{0}
\renewcommand\thetheorem{S\arabic{theorem}}

Recall that for $z=0,1$, $B_i(z)=Y_i(z)-\tau_{CACE} W_i(z)$. For $z=0,1$, let $\overline{B}(z)$ be the mean of $B_i(z)$ for all $n$ units, and let $\mathbb{S}_{B(z)}^2$ denote the finite population variance of $B_i(z)$. Extending the conditions in \cite{li2017general}, we introduce Condition~\ref{lindeberg-feller} to study the asymptotic properties of the Wald estimator in Theorems 1 and 2.
\begin{condition}
	\label{lindeberg-feller}
	As $n\rightarrow\infty$, (1) the limit inferior of $p_{co}$ is positive; (2) the proportion of units under treatment, $n_1/n$, has a limit in $(0,1)$; (3) the limit superiors of $\mathbb{S}_{B(0)}^2$ and $\mathbb{S}_{B(1)}^2$ are finite; (4) for $z=0,1$, the following Lindeberg-Feller type condition holds:
	\begin{equation}
		\frac{\max_{1 \leq i \leq n}\left(B_{i}(z)-\overline{B}(z)\right)^2}{\min(n_1, n_0)^2\cdot Var\left(\widehat{\tau}_B\right)}\rightarrow0.\label{eq:lindeberg-feller}
	\end{equation}
\end{condition}

Let $\mathbb{S}_{B(1)-B(0)}^2$ denote the finite population variance of $B_i(1)-B_i(0)$. Let $\mathbb{S}_{\bm{x}\bm{x}}$ denote the finite population covariance matrix of the covariates,
\begin{equation*}
	\mathbb{S}_{\bm{x}\bm{x}}=\frac{1}{n-1}\sum_{i=1}^{n}(\bm{x}_{i}-\overline{\bm{x}})(%
	\bm{x}_{i}-\overline{\bm{x}})^{\top }=\frac{1}{n-1}\sum_{i=1}^{n}\bm{x}_{i}\bm{x}_{i}^{\top}. \label{eq:Cov}
\end{equation*}
\noindent For $z=0,1$, let $\mathbb{S}_{B(z),\bm{x}}$ be the finite population covariance vector between $B_i(z)$ and $\bm{x}_i$,
\begin{equation*}
	\mathbb{S}_{B(z),\bm{x}}=\frac{1}{n-1}\sum_{i=1}^{n}(B_i(z)-\overline{B}(z))(\bm{x}_i-\overline{\bm{x}}).
\end{equation*}
Let $\left\|\bm{x}_{i}\right\|$ denote the Euclidean distance between $\bm{x}_i$ and $\overline{\bm{x}}=\bm{0}$. Extending the conditions in \cite{li2020rerandomization}, we introduce Condition~\ref{strict_cond} to study the asymptotic properties of the regression adjustment estimator in Theorems 3-5.
\begin{condition}
	\label{strict_cond}
	As $n\rightarrow\infty$, (1) the limit inferior of $p_{co}$ is positive; (2) the proportion of units under treatment, $n_1/n$, has a limit in $(0,1)$; (3) $\mathbb{S}_{B(z)}^2$ ($z=0,1$), $\mathbb{S}_{B(1)-B(0)}^2$, $\mathbb{S}_{\bm{x}\bm{x}}$ and $\mathbb{S}_{B(z),\bm{x}}$ ($z=0,1$) have finite limiting values, and the limit of $\mathbb{S}_{\bm{x}\bm{x}}$ is non-singular; (4) $\max_{1\leq i\leq n}\left|B_i(z)-\overline{B}(z)\right|^{2} / n \rightarrow 0$ ($z=0,1$) and $\max_{1\leq i\leq n}\left\|\bm{x}_{i}\right\|^{2}/n$ $\rightarrow 0$.
\end{condition}

\noindent{\bf Remark.} With (4), we have
\begin{equation*}
\begin{aligned}
	&\frac{\max_{1 \leq i \leq n}\left(B_{i}(z)-\overline{B}(z)\right)^2}{\min(n_1, n_0)^2\cdot Var\left(\widehat{\tau}_B\right)}\\
=&\frac{n^2}{\min(n_1, n_0)^2}\cdot\frac{\max_{1\leq i\leq n}\left|B_i(z)-\overline{B}(z)\right|^{2}/n}{n\left(\mathbb{S}_{B(1)}/n_1+\mathbb{S}_{B(0)}/n_0-\mathbb{S}_{B(1)-B(0)}/n\right)}\\
=&\left(\frac{n}{\min(n_1, n_0)}\right)^2\cdot\frac{\max_{1\leq i\leq n}\left|B_i(z)-\overline{B}(z)\right|^{2}/n}{n/n_1\cdot\mathbb{S}_{B(1)}+n/n_0\cdot\mathbb{S}_{B(0)}-\mathbb{S}_{B(1)-B(0)}}.
\end{aligned}
\end{equation*}
Under assumption (2) of Condition~\ref{strict_cond}, as $n\rightarrow\infty$, $n/\min(n_1,n_0)$, $n/n_1$ and $n/n_0$ all have a finite limit.  Together with assumptions (3) and (4) of Condition \ref{strict_cond}, we have \eqref{eq:lindeberg-feller}. Also, assumptions (1)-(3) of Condition~\ref{strict_cond} imply assumptions (1)-(3) of Condition~\ref{lindeberg-feller}. Hence, Condition~\ref{strict_cond} implies Condition~\ref{lindeberg-feller}, and conclusions obtained under Condition~\ref{lindeberg-feller} automatically hold under Condition~\ref{strict_cond}.

\section*{S.1 Propositions and Their Proof}

For any set of potential outcomes $\{(Q_i(0),Q_i(1)):\ i=1,\cdots,n\}$, let $Q_i=Q_i(Z_i)$ be the realized outcome under the assigned treatment arm.  For units with $Z_i=z$ ($z=0,1$), let $\overline{Q}_z$ denote the sample mean of $Q_i=Q_i(z)$: $\overline{Q}_z=\sum_{i:\ Z_i=z}Q_i/n_z$, and let $S_{Q(z)}^2$ denote the sample variance of $Q_i=Q_i(z)$:
\begin{equation*}
	S_{Q(z)}^2=\frac{1}{n_z-1}\sum_{Z_i=z}\left(Q_i-\overline{Q}_z\right)^2.
\end{equation*}
Also, for any two sets of potential outcomes $\{(Q_i(0),Q_i(1)):\ i=1,\cdots,n\}$ and $\{(Q_i'(0),Q_i'(1)):\ i=1,\cdots,n\}$, for units with $Z_i=z$ ($z=0,1$), let $S_{Q(z),Q'(z)}$ denote the sample covariance between $Q_i=Q_i(z)$ and $Q'_i=Q'_i(z)$:
\begin{equation*}
	S_{Q(z),Q'(z)}=\frac{1}{n_z-1}\sum_{Z_i=z}\left(Q_i-\overline{Q}_z\right)\left(Q_i'-\overline{Q}'_z\right).
\end{equation*}
Finally, for units with $Z_i=z$ ($z=0,1$), let $\bm{S}_{Q(z),\bm{x}}$ ($z=0,1$) denote the vector of sample covariances between $Q_i=Q_i(z)$ and $\bm{x}_i$:
\begin{equation*}	\bm{S}_{Q(z),\bm{x}}=\frac{1}{n_z-1}\sum_{Z_i=z}\left(Q_i-\overline{Q}_z\right)\left(\bm{x}_i-\overline{\bm{x}}_z\right)^{\top},
\end{equation*}
where $\overline{\bm{x}}_z=\sum_{i:\ Z_i=z}\bm{x}_i/n_z$.

\setcounter{theorem}{0}
\begin{proposition}
	The confidence interval in (8),
$$\widehat{\tau}^{Wald}\pm \nu_{1-\alpha/2}\sqrt{\widehat{Var}(\widehat{\tau}_B)}/\widehat{\tau}_W,$$
is the same as the super-population confidence interval for $\tau_{CACE}^{pop}$ obtained by the delta method.
\end{proposition}
\subsection*{S.1.1 Proof of Proposition S1:}
For units with $Z_i=z$ ($z=0,1$), let $S_{Y(z)}^2$ be the sample variance of $Y_i$, $S_{W(z)}^2$ the sample variance of $W_i$, and $S_{Y(z),W(z)}$ the sample covariance between $Y_i$ and $W_i$. Because $\widehat{B}_i=Y_i-\widehat{\tau}^{Wald}W_i=Y_i-(\widehat{\tau}_Y/\widehat{\tau}_W)W_i$, we have
\begin{equation*}
S_{\widehat{B}(z)}^2=S_{Y(z)}^2+\frac{\widehat{\tau}_Y^2}{\widehat{\tau}_W^2}S_{W(z)}^2-2\frac{\widehat{\tau}_Y}{\widehat{\tau}_W}S(Y(z),W(z)).
\end{equation*}
Combining this with (7), we have
\begin{equation*}
\begin{aligned}
\widehat{Var}(\widehat{\tau}_B)=&\left(\frac{S_{Y(1)}^2}{n_1}+\frac{S_{Y(0)}^2}{n_0}\right)+\frac{\widehat{\tau}_Y^2}{\widehat{\tau}_W^2}\left(\frac{S_{W(1)}^2}{n_1}+\frac{S_{W(0)}^2}{n_0}\right)\\
&-2\frac{\widehat{\tau}_Y}{\widehat{\tau}_W}\left(\frac{S_{Y(1),W(1)}}{n_1}+\frac{S_{Y(0),W(0)}}{n_0}\right).
\end{aligned}
\end{equation*}

To construct confidence interval for $\tau_{CACE}^{pop}$, according to the delta method (see e.g. \cite{imbens2015causal}, Ch. 23), in large samples, $\widehat{\tau}^{Wald}$ is approximately normally distributed with mean $\tau_{CACE}^{pop}$ and variance
\begin{equation*}
Var_{sp}\left(\widehat{\tau}^{Wald}\right)=\frac{1}{\tau_W^2}\left(Var_{sp}(\widehat{\tau}_Y)+\frac{\tau_Y^2}{\tau_W^2}Var_{sp}(\widehat{\tau}_W)-2\frac{\tau_Y}{\tau_W}Cov_{sp}(\widehat{\tau}_Y,\widehat{\tau}_W)\right),
\end{equation*}
where the subscript ``sp" indicates that the variances and covariance are for super-population. Since units are assumed to be independently and randomly sampled from the super-population, it is easy to see that the estimate for the super-population variance of $\widehat{\tau}_Y$ is
\begin{equation*}
\widehat{Var}_{sp}(\widehat{\tau}_Y)=\widehat{Var}_{sp}(\overline{Y}_1-\overline{Y}_0)=\frac{S_{Y(1)}^2}{n_1}+\frac{S_{Y(0)}^2}{n_0}.
\end{equation*}
Similarly, the estimates for the super-population variance of $\widehat{\tau}_W$ and the super-population covariance between $\widehat{\tau}_Y$ and $\widehat{\tau}_W$ are
\begin{equation*}
\begin{aligned}
\widehat{Var}_{sp}(\widehat{\tau}_W)&=\frac{S_{W(1)}^2}{n_1}+\frac{S_{W(0)}^2}{n_0},\\
Cov_{sp}(\widehat{\tau}_Y,\widehat{\tau}_W)&=\frac{S_{Y(1),W(1)}}{n_1}+\frac{S_{Y(0),W(0)}}{n_0}.
\end{aligned}
\end{equation*}
Further with $\widehat{\tau}_Y$ and $\widehat{\tau}_W$ as the estimates for $\tau_Y$ and $\tau_W$, it is easy to see that the estimate of $Var_{sp}\left(\widehat{\tau}^{Wald}\right)$ can be written as $\widehat{Var}_{sp}\left(\widehat{\tau}^{Wald}\right)=\widehat{Var}(\widehat{\tau}_B)/\widehat{\tau}_W^2$.
The super-population confidence interval for $\tau_{CACE}^{pop}$, $\widehat{\tau}^{Wald}\pm\nu_{1-\alpha/2}\sqrt{\widehat{Var}_{sp}\left(\widehat{\tau}^{Wald}\right)}$, is thus the same as that in (8).

\begin{proposition}
	For any set of quantities $\{Q_i:\ i=1\ldots,n\}$, the coefficient on $Z_i$ (denoted by $\widehat{\tau}_Q^{Reg}$) in the linear projection of $Q_i$ on $Z_i$, $\bm{x}_i$ and $Z_i\bm{x}_i$ using all $n$ units equals to
	\begin{equation*}
		\label{itt_adjust_general}
		\frac{1}{n_1}\sum_{i:\ Z_i=1}\left(Q_i-\widehat{\bm{\beta}}_{Q,1}^{\top}\bm{x}_i\right)-\frac{1}{n_0}\sum_{i:\ Z_i=0}\left(Q_i-\widehat{\bm{\beta}}_{Q,0}^{\top}\bm{x}_i\right).
	\end{equation*}
\end{proposition}

\subsection*{S.1.2 Proof of Proposition S2:}
In the linear projection of $Q_i$ on $Z_i$, $\bm{x}_i$ and $Z_i\bm{x}_i$, the intercept and the slope coefficients are
\begin{equation}\label{prop_A2_eq1}
	\begin{aligned} &\left(\widehat{\beta}_0,\widehat{\tau}_Q^{Reg},\widehat{\bm{\beta}}_{\bm{x}},\widehat{\bm{\beta}}_{Z\bm{x}}\right)\\
=&\text{argmin}_{\beta_0,\tau_Q^{Reg},\bm{\beta}_{\bm{x}},\bm{\beta}_{Z\bm{x}} }\sum_{i=1}^n\left(Q_i-\beta_0-\tau_Q^{Reg}Z_i-\bm{\beta}_{\bm{x}}^{\top}\bm{x}_i-\bm{\beta}_{Z\bm{x}}^{\top}(Z_i\bm{x}_i)\right)^2\\ =&\text{argmin}_{\beta_0,\tau_Q^{Reg},\bm{\beta}_{\bm{x}},\bm{\beta}_{Z\bm{x}} }\left(\sum_{i:\ Z_i=0}\left(Q_i-\beta_0-\bm{\beta}_{\bm{x}}^{\top}\bm{x}_i\right)^2+\right.\\
&\left.\sum_{i:\ Z_i=1}\left(Q_i-\beta_0-\tau_{Q}^{Reg}-\bm{\beta}_{\bm{x}}^{\top}\bm{x}_i-\bm{\beta}_{Z\bm{x}}^{\top}\bm{x}_i\right)^2\right)
	\end{aligned}
\end{equation}

In order to minimize the part in the big parenthesis in \eqref{prop_A2_eq1}, we can take the following two steps: (1) find $\widehat{\beta}_0$ and $\widehat{\bm{\beta}}_{\bm{x}}$ that minimize the first term in the big parenthesis, $\sum_{i:\ Z_i=0}\left(Q_i-\beta_0-\bm{\beta}_{\bm{x}}^{\top}\bm{x}_i\right)^2$; (2) fix $\widehat{\beta}_0$ and $\widehat{\bm{\beta}}_{\bm{x}}$, and find $\widehat{\tau}_Q^{Reg}$ and $\widehat{\bm{\beta}}_{Z\bm{x}}$ that minimize the second term in the big parenthesis,\\ $\sum_{i:Z_i=1}\left(Q_i-\widehat{\beta}_0-\tau_Q^{Reg}-\widehat{\bm{\beta}}_{\bm{x}}^{\top}\bm{x}_i-\bm{\beta}_{Z\bm{x}}^{\top}\bm{x}_i\right)^2$.
From step 1, it is easy to see that $\widehat{\bm{\beta}}_{Q,0}=\widehat{\bm{\beta}}_{\bm{x}}$.  Step 2 is equivalent to finding $\widehat{\alpha}_{Q,1}$ and $\widehat{\bm{\beta}}_{Q,1}$ that minimize $\sum_{i:Z_i=1}\left(Q_i-\alpha_{Q,1}-\bm{\beta}_{Q,1}^{\top}\bm{x}_i\right)^2$, with $\widehat{\alpha}_{Q,1}=\widehat{\beta}_0+\widehat{\tau}_Q^{Reg}$ and  $\widehat{\bm{\beta}}_{Q,1}=\widehat{\bm{\beta}}_{\bm{x}}+\widehat{\bm{\beta}}_{Z\bm{x}}$.

With the linear projection, we can write
\begin{equation*}
Q_i=\widehat{\beta}_0+\widehat{\tau}_Q^{Reg}Z_i+\widehat{\bm{\beta}}_{\bm{x}}^{\top}\bm{x}_i+\widehat{\bm{\beta}}_{Z\bm{x}}^{\top}(Z_i\bm{x}_i)+\widehat{\varepsilon}_{Q,i}.
\end{equation*}
For units with $Z_i=0$, $\widehat{\varepsilon}_{Q,i}=Q_i-\widehat{\beta}_0-\widehat{\bm{\beta}}_{\bm{x}}^{\top}\bm{x}_i$; for units with $Z_i=1$, $\widehat{\varepsilon}_{Q,i}=Q_i-\widehat{\beta}_0-\widehat{\tau}_Q^{Reg}-\widehat{\bm{\beta}}_{\bm{x}}^{\top}\bm{x}_i-\widehat{\bm{\beta}}_{Z\bm{x}}\bm{x}_i$.
Due to properties of the linear projection, $\widehat{\varepsilon}_{Q,i}$ is orthogonal to the constant 1, and also orthogonal to $Z_i$. Therefore $\sum_{i=1}^n \widehat{\varepsilon}_{Q,i}=0$ and $\sum_{i=1}^n Z_i\widehat{\varepsilon}_{Q,i}=0$. Because $\sum_{i=1}^n \widehat{\varepsilon}_{Q,i}=\sum_{i:\ Z_i=0} \widehat{\varepsilon}_{Q,i}+\sum_{i:\ Z_i=1} \widehat{\varepsilon}_{Q,i}$ and $\sum_{i=1}^n Z_i\widehat{\varepsilon}_{Q,i}=\sum_{i:\ Z_i=1}\widehat{\varepsilon}_{Q,i}$, we have $\sum_{i:\ Z_i=0}\widehat{\varepsilon}_{Q,i}=\sum_{i:\ Z_i=1}\widehat{\varepsilon}_{Q,i}=0$.

Plugging in the expressions of $\widehat{\varepsilon}_{Q,i}$ for units with $Z_i=0$ and $Z_i=1$, we have
\begin{equation*}
\begin{aligned}
&\sum_{i:\ Z_i=0}\left(Q_i-\widehat{\beta}_0-\widehat{\bm{\beta}}_{\bm{x}}^{\top}\bm{x}_i\right)=0\\
\implies&\widehat{\beta}_0=\frac{1}{n_0}\sum_{i:\ Z_i=0}\left(Q_i-\widehat{\bm{\beta}}_{\bm{x}}^{\top}\bm{x}_i\right)=\frac{1}{n_0}\sum_{i:\ Z_i=0}\left(Q_i-\widehat{\bm{\beta}}_{Q,0}^{\top}\bm{x}_i\right)
\end{aligned}
\end{equation*}
and
\begin{equation*}
\begin{aligned}
&\sum_{i:\ Z_i=1}\left(Q_i-\widehat{\beta}_0-\widehat{\tau}_Q^{Reg}-\widehat{\bm{\beta}}_{\bm{x}}^{\top}\bm{x}_i-\widehat{\bm{\beta}}_{Z\bm{x}}^{\top}\bm{x}_i\right)=0\\
\implies&\widehat{\beta}_0+\widehat{\tau}_Q^{Reg}=\frac{1}{n_1}\sum_{i:\ Z_i=1}\left(Q_i-\widehat{\bm{\beta}}_{\bm{x}}^{\top}\bm{x}_i-\widehat{\bm{\beta}}_{Z\bm{x}}^{\top}\bm{x}_i\right)=\frac{1}{n_1}\sum_{i:\ Z_i=1}\left(Q_i-\widehat{\bm{\beta}}_{Q,1}^{\top}\bm{x}_i\right)
\end{aligned}
\end{equation*}
Hence $\widehat{\tau}_Q^{Reg}=\frac{1}{n_1}\sum_{i:\ Z_i=1}\left(Q_i-\widehat{\bm{\beta}}_{Q,1}^{\top}\bm{x}_i\right)-\frac{1}{n_0}\sum_{i:\ Z_i=0}\left(Q_i-\widehat{\bm{\beta}}_{Q,0}^{\top}\bm{x}_i\right)$.

\section*{S.2 Proof of Theorems}
\subsection*{S.2.1 Proof of Theorem 1}

Let $\mathbb{S}^2_{W(z)}$ ($z=0,1$) denote the finite population variance of $W_i(z)$, which is bounded since $W_i(z)$ is binary. Let $\mathbb{S}^2_{W(1)-W(0)}$ denote the finite population variance of $W_i(1)-W_i(0)$. Applying Theorem 3 in \cite{li2017general}, over all randomizations, $\widehat{\tau}_W$ has mean $\tau_W=p_{co}$ and variance $Var(\widehat{\tau}_W)=\mathbb{S}^2_{W(1)}/n_1+\mathbb{S}^2_{W(0)}/n_0-\mathbb{S}^2_{W(1)-W(0)}/n\leq\mathbb{S}^2_{W}(1)/n_1+\mathbb{S}^2_{W}(0)/n_0$.
Since $n_1/n$ has a limit in $(0,1)$, we have $\lim_{n\rightarrow\infty} 1/n_1=\lim_{n\rightarrow\infty}1/n_0=0$. By Chebyshev's inequality, for any $\epsilon>0$,
\begin{align*}
	&\lim_{n\rightarrow\infty}Pr\left(|\widehat{\tau}_W-p_{co}|\geq\epsilon\right)\\ \leq&\lim_{n\rightarrow\infty}\frac{Var(\widehat{\tau}_W)}{\epsilon^2}\leq\lim_{n\rightarrow\infty}\frac{1}{\epsilon^2}\left(\frac{\mathbb{S}^2_{W(1)}}{n_1}+\frac{\mathbb{S}^2_{W(0)}}{n_0}\right)=0,
\end{align*}
that is, $\widehat{\tau}_W-p_{co}=o_p(1)$.
Similarly, given that the limit superiors of $\mathbb{S}^2_{B(0)}$ and $\mathbb{S}^2_{B(1)}$ are finite, we have $\widehat{\tau}_B-\tau_B=\widehat{\tau}_B=o_p(1)$ (noting that $\tau_B=0$ from Section~3.2.2).

Combining (2) and (3), we have
\begin{equation}
	\frac{\widehat{\tau}_B}{\widehat{\tau}_W}=\widehat{\tau}^{Wald}-\tau_{CACE}.\label{diff_Wald}
\end{equation}
Since $p_{co}$ has a positive limit inferior, $\widehat{\tau}_W$ has a positive limit inferior in probability. Hence $\widehat{\tau}_B/\widehat{\tau}_W=o_p(1)$, which implies that $\widehat{\tau}^{Wald}-\tau_{CACE}=o_p(1)$.

\subsection*{S.2.2 Proof of Theorem 2}
For $z=0,1$, define
\begin{equation*}
	Var(\widehat{\tau}_B)^{+}=\frac{\mathbb{S}_{B(1)}^2}{n_1}+\frac{\mathbb{S}_{B(0)}^2}{n_0},
\end{equation*}
and also define
\begin{equation*}
	\label{estimated_SA}
	\widetilde{Var}(\widehat{\tau}_B)=\frac{S_{B(1)}^2}{n_1}+\frac{S_{B(0)}^2}{n_0},
\end{equation*}
where $S_{B(z)}^2$ is the sample variance of $B_i=B_i(z)$ for units with $Z_i=z$. We first prove that $|\widehat{Var}(\widehat{\tau}_B)-\widetilde{Var}(\widehat{\tau}_B)|/Var\left(\widehat{\tau}_B\right)^{+}=o_p(1)$ as $n\rightarrow\infty$. (Recall that $\widehat{Var}(\widehat{\tau}_B)=S_{\widehat{B}(1)}^2/n_1+S_{\widehat{B}(0)}^2/n_0$.)

According to the expressions of $B_i$ and $\widehat{B}_i$,  $\widehat{B}_i=B_i-\left(\widehat{\tau}^{Wald}-\tau_{CACE}\right)W_i$. Let $S_{W(z)}^2$ be sample variance of $W_i$ for units with $Z_i=z$, and let $S_{B(z),W(z)}$ be the sample covariance between $B_i$ and $W_i$ for units with $Z_i=z$. We have
\begin{align*}
S_{\widehat{B}(z)}^2=S_{B(z)}^2-2\left(\widehat{\tau}^{Wald}-\tau_{CACE}\right)S_{B(z),W(z)}+\left(\widehat{\tau}^{Wald}-\tau_{CACE}\right)^2S_{W(z)}^2
\end{align*}
for $z=0,1$.
The absolute difference between $\widehat{Var}(\widehat{\tau}_B)$ and $\widetilde{Var}(\widehat{\tau}_B)$ can be bounded as follows:
\begin{equation}
	\begin{aligned}
		&\left|\widehat{Var}(\widehat{\tau}_B)-\widetilde{Var}(\widehat{\tau}_B)\right|\\
		=&\left|\frac{-2\left(\widehat{\tau}^{Wald}-\tau_{CACE}\right)S_{B(1),W(1)}
			+\left(\widehat{\tau}^{Wald}-\tau_{CACE}\right)^2S^2_{W(1)}}{n_1}\right.\\
		&\left.+\frac{-2\left(\widehat{\tau}^{Wald}-\tau_{CACE}\right)S_{B(0),W(0)}
			+\left(\widehat{\tau}^{Wald}-\tau_{CACE}\right)^2S^2_{W(0)}}{n_0}\right|\\ \leq&\frac{2\left|\widehat{\tau}^{Wald}-\tau_{CACE}\right|\left|S_{B(1),W(1)}\right|+\left(\widehat{\tau}^{Wald}-\tau_{CACE}\right)^2S^2_{W(1)}}{n_1}\\
		&+\frac{2\left|\widehat{\tau}^{Wald}-\tau_{CACE}\right|\left|S_{B(0),W(0)}\right|+
			\left(\widehat{\tau}^{Wald}-\tau_{CACE}\right)^2S^2_{W(0)}}{n_0}.
	\end{aligned}\label{abs_diff}
\end{equation}

Because $W_i$ is binary, we have
\begin{equation}
	S_{W(z)}^2=\frac{1}{n_z-1}\sum_{i:\ Z_i=z}(W_i-\overline{W}_z)^2\leq\frac{n_z}{n_z-1}\label{S_Wz_leq}
\end{equation}
and
\begin{equation}
	\begin{aligned}
		\left|S_{B(z), W(z)}\right|&=\left|\frac{1}{n_z-1}\sum_{i:\ Z_i=z}(B_i-\overline{B}_z)(W_i-\overline{W}_z)\right|\\
&\leq\frac{n_z}{n_z-1}\max_{i:Z_i=z}\left|B_{i}(z)-\overline{B}_z\right|\\
&\leq\frac{n_z}{n_z-1}\left(\max_{i:Z_i=z}\left|B_{i}(z)-\overline{B}(z)\right|+\left|\overline{B}(z)-\overline{B}_z\right|\right)\\
&=\frac{n_z}{n_z-1}\left(\max_{i:Z_i=z}\left|B_{i}(z)-\overline{B}(z)\right|+\left|\overline{B}(z)-\frac{1}{n_z}\sum_{i:\ Z_i=z}B_i(z)\right|\right)\\
		&\leq\frac{n_z}{n_z-1}\left(\max_{i:Z_i=z}\left|B_{i}(z)-\overline{B}(z)\right|+\frac{1}{n_z}\sum_{i:Z_{i}=z}\left|\overline{B}(z)-{B}_i(z)\right|\right)\\
		&\leq\frac{2n_z}{n_z-1}\max_{i:Z_i=z}\left|B_{i}(z)-\overline{B}(z)\right|\\&\leq\frac{2n_z}{n_z-1}\max_{1\leq{}i\leq{}n}\left|B_{i}(z)-\overline{B}(z)\right|\label{S_AzWz_leq}
	\end{aligned}
\end{equation}
for $z=0,1$.
Plugging \eqref{S_Wz_leq} and \eqref{S_AzWz_leq} into \eqref{abs_diff}, we have
\begin{equation}
	\begin{aligned}
		&\left|\widehat{Var}(\widehat{\tau}_B)-\widetilde{Var}(\widehat{\tau}_B)\right|\\
		\leq&\frac{4\left|\widehat{\tau}^{Wald}-\tau_{CACE}\right|\max_{1 \leq i \leq n}\left|B_{i}(1)-\overline{B}(1)\right|}{n_1-1}+\frac{\left(\widehat{\tau}^{Wald}-\tau_{CACE}\right)^2}{n_1-1}\\
		&+\frac{4\left|\widehat{\tau}^{Wald}-\tau_{CACE}\right|\max_{1 \leq i \leq n}\left|B_{i}(0)-\overline{B}(0)\right|}{n_0-1}+\frac{\left(\widehat{\tau}^{Wald}-\tau_{CACE}\right)^2}{n_0-1}.
	\end{aligned}\label{abs_diff2}
\end{equation}
Substituting \eqref{diff_Wald} into \eqref{abs_diff2} and dividing by $Var\left(\widehat{\tau}_B\right)^{+}$, we have
\begin{equation}
	\begin{aligned}
		&\left|\widehat{Var}(\widehat{\tau}_B)-\widetilde{Var}(\widehat{\tau}_B)\right|/Var\left(\widehat{\tau}_B\right)^{+}\\
		\leq&\frac{4\max_{1 \leq i \leq n}\left|B_{i}(1)-\overline{B}(1)\right|}{(n_1-1)\sqrt{Var\left(\widehat{\tau}_B\right)^{+}}}\times|\widehat{\tau}_{W}^{-1}|\times\left|\frac{\widehat{\tau}_B}{\sqrt{Var\left(\widehat{\tau}_B\right)^{+}}}\right|\\
		&+\frac{\widehat{\tau}_{W}^{-2}}{n_1-1}\left(\frac{\widehat{\tau}_B}{\sqrt{Var\left(\widehat{\tau}_B\right)^{+}}}\right)^2\\
		&+\frac{4\max_{1 \leq i \leq n}\left|B_{i}(0)-\overline{B}(0)\right|}{(n_0-1)\sqrt{Var\left(\widehat{\tau}_B\right)^{+}}}\times|\widehat{\tau}_{W}^{-1}|\times\left|\frac{\widehat{\tau}_B}{\sqrt{Var\left(\widehat{\tau}_B\right)^{+}}}\right|\\
		&+\frac{\widehat{\tau}_{W}^{-2}}{n_0-1}\left(\frac{\widehat{\tau}_B}{\sqrt{Var\left(\widehat{\tau}_B\right)^+}}\right)^2\\
		\leq&\frac{8\max_{1 \leq i \leq n}|B_{i}(1)-\overline{B}(1)|}{\min(n_1, n_0)\sqrt{Var\left(\widehat{\tau}_B\right)}}\times|\widehat{\tau}_{W}^{-1}|\times\left|\frac{\widehat{\tau}_B}{\sqrt{Var\left(\widehat{\tau}_B\right)}}\right|\\
		&+\frac{8\max_{1 \leq i \leq n}|B_{i}(0)-\overline{B}(0)|}{\min(n_1, n_0)\sqrt{Var\left(\widehat{\tau}_B\right)}}\times|\widehat{\tau}_{W}^{-1}|\times\left|\frac{\widehat{\tau}_B}{\sqrt{Var\left(\widehat{\tau}_B\right)}}\right|\\
		&+\frac{4\widehat{\tau}_{W}^{-2}}{\min(n_1,n_0)}\left(\frac{\widehat{\tau}_B}{\sqrt{Var\left(\widehat{\tau}_B\right)}}\right)^2,
	\end{aligned}\label{abs_diff3}
\end{equation}
where the last inequality is derived using the following inequalities:
\begin{equation*}
\begin{aligned}
&\frac{1}{n_z-1}\leq \frac{2}{n_z}\leq \frac{2}{\min(n_0,n_1)},\quad z=0,1,\\
&Var\left(\widehat{\tau}_B\right)^{+}\geq Var\left(\widehat{\tau}_B\right).
\end{aligned}
\end{equation*}
(Assumption (2) in Condition \ref{lindeberg-feller} implies that $n_z\rightarrow \infty$ as $n\rightarrow\infty$, and hence we can assume $n_z\geq 2$, which implies that $\frac{1}{n_z-1}\leq \frac{2}{n_z}$.)

Equation~\eqref{eq:lindeberg-feller} in Condition \ref{lindeberg-feller} implies that
\begin{equation*}
	\lim_{n\rightarrow\infty}\frac{\max_{1 \leq i \leq n}\left|B_{i}(z)-\overline{B}(z)\right|}{\min(n_1, n_0)\cdot \sqrt{Var\left(\widehat{\tau}_B\right)}}=0,\quad\text{for }z=0,1.
\end{equation*}
According to assumption (1) in Condition \ref{lindeberg-feller}, the limit inferior of $p_{co}$ is positive. Since $\widehat{\tau}_{W}$ is a consistent estimator of $p_{co}$, $\widehat{\tau}_{W}$ has a positive limit inferior in probability, and therefore $\widehat{\tau}_{W}^{-1}$ has a positive limit superior in probability.  Assumption (2) in Condition \ref{lindeberg-feller} implies that $\min(n_1,n_0)\rightarrow \infty$ as $n\rightarrow\infty$.  Under Condition \ref{lindeberg-feller}, we can apply Theorem 4 of \cite{li2017general} to have $\widehat{\tau}_B/\sqrt{Var\left(\widehat{\tau}_B\right)}\stackrel{d}{\longrightarrow}N(0,1)$ as $n\rightarrow\infty$. We then apply Slutsky's theorem and the continuous mapping theorem to conclude that the right hand side of \eqref{abs_diff3} converges to 0 in probability.  Hence $|\widehat{Var}(\widehat{\tau}_B)-\widetilde{Var}(\widehat{\tau}_B)|/Var\left(\widehat{\tau}_B\right)^{+}\stackrel{p}{\longrightarrow}0$ as $n\rightarrow \infty$.

Under Condition \ref{lindeberg-feller}, we can apply Proposition 1 of \cite{li2017general} to have \\ $\widetilde{Var}(\widehat{\tau}_B)/Var\left(\widehat{\tau}_B\right)^{+}\stackrel{p}{\longrightarrow}1$ as $n\rightarrow\infty$. It follows that $\widehat{Var}(\widehat{\tau}_B)/Var(\widehat{\tau}_B)^{+}$ $\stackrel{p}{\longrightarrow}1$ as $n\rightarrow\infty$. Because $Var\left(\widehat{\tau}_B\right)^{+}\geq Var\left(\widehat{\tau}_B\right)$, $\widehat{Var}(\widehat{\tau}_B)$ is an asymptotically conservative estimator of $Var\left(\widehat{\tau}_B\right)$.  Combining this with the statement that (6) has $(1-\alpha)$ coverage rate for $\tau_{CACE}$ as $n\rightarrow\infty$, Theorem 2 is proved.

\subsection*{S.2.3 Proof of Theorem 3}
For $z=0,1$, for units with $Z_i=z$, let
\begin{equation*}
	\bm{S}_{\bm{x}\bm{x},z}=\frac{1}{n_z-1}\sum_{i:Z_{i}=z}(\bm{x}_i-\overline{\bm{x}}_z)(\bm{x}_i-\overline{\bm{x}}_z)^{\top }
\end{equation*}
denote the sample covariance matrix for the covariates, with $\overline{\bm{x}}_z=\frac{1}{n_z}\sum_{i:Z_{i}=z}\bm{x}_i$ being the mean of covariates for units with $Z_i=z$.
Let $\bm{S}_{\bm{x}\bm{x},z}=\bm{C}^{\top}\bm{C}$ be its Cholesky decomposition. Note that $\widehat{\bm{\beta}}_{W,z}$ is the vector of coefficients on $\bm{x}$ in the linear projection of $W_i=W_i(z)$ on $\bm{x}_i$ using units with $Z_i=z$. For units with $Z_i=z$, the sample variance of the linear projection $\widehat{\bm{\beta}}_{W,z}^{\top}\bm{x}$, which equals $\widehat{\bm{\beta}}_{W,z}^{\top}\bm{S}_{\bm{x}\bm{x},z}\widehat{\bm{\beta}}_{W,z}=\widehat{\bm{\beta}}_{W,z}^{\top}\bm{C}^{\top}\bm{C}\widehat{\bm{\beta}}_{W,z}$, is no larger than the sample variance of $W_i=W_i(z)$, $S^2_{W(z)}$. Hence
\begin{equation}
	\label{bound_on_projection}
	\begin{aligned} &\left(\widehat{\bm{\beta}}_{W,z}^{\top}\overline{\bm{x}}_z\right)^2=\left(\left(\widehat{\bm{\beta}}_{W,z}^{\top}\bm{C}^{\top}\right)\left(\bm{C}\bm{S}_{\bm{x}\bm{x},z}^{-1}\overline{\bm{x}}_z\right)\right)^2\\
		\leq&\left(\widehat{\bm{\beta}}_{W,z}^{\top}\bm{C}^{\top}\bm{C}\widehat{\bm{\beta}}_{W,z}\right)\left(\overline{\bm{x}}_z^{\top}\bm{S}_{\bm{x}\bm{x},z}^{-1}\bm{C}^{\top}\bm{C}\bm{S}_{\bm{x}\bm{x},z}^{-1}\overline{\bm{x}}_z\right)\\
		\leq&\ S^2_{W(z)}\left(\overline{\bm{x}}_z^{\top}\bm{S}_{\bm{x}\bm{x},z}^{-1}\overline{\bm{x}}_z\right)\leq\frac{n_z}{n_z-1}\overline{\bm{x}}_z^{\top}\bm{S}_{\bm{x}\bm{x},z}^{-1}\overline{\bm{x}}_z,
	\end{aligned}
\end{equation}
where the first inequality is obtained using Cauchy-Schwarz inequality and the last is obtained using \eqref{S_Wz_leq}.

Let the $k$th covariate of unit $i$ be $x_i^{(k)}$, let $\overline{x}_z^{(k)}$ ($z=0,1$) denote the mean of $x_i^{(k)}$ for units with $Z_i=z$, and let $\mathbb{S}_{\bm{x}\bm{x}}^{(k,k)}$ denote the $k$th diagonal element of $\mathbb{S}_{\bm{x}\bm{x}}$, that is, the finite population variance of $x_i^{(k)}$.  We have
\begin{equation*}
\label{eq:lindeberg-feller-x}
\begin{aligned}
\frac{\max_{1\leq i\leq n}\left(x_i^{(k)}\right)^2}{\min(n_1,n_0)\mathbb{S}_{\bm{x}\bm{x}}^{(k,k)}}&=\frac{\max_{1\leq i\leq n}\left(x_i^{(k)}\right)^2}{n}\cdot\frac{n}{\min(n_1,n_0)}\cdot\frac{1}{\mathbb{S}_{\bm{x}\bm{x}}^{(k,k)}}\\
&\leq \frac{\max_{1\leq i\leq n}\left\|\bm{x}_{i}\right\|^2}{n}\cdot\frac{n}{\min(n_1,n_0)}\cdot\frac{1}{\mathbb{S}_{\bm{x}\bm{x}}^{(k,k)}}.
\end{aligned}
\end{equation*}
Under assumption (4) of Condition \ref{strict_cond}, $\max_{1\leq i\leq n}\left\|\bm{x}_{i}\right\|^2/n\rightarrow 0$; under assumption (2) of Condition \ref{strict_cond}, $n/\min(n_1,n_0)$ has a finite limiting value; under assumption (3) of Condition \ref{strict_cond}, $1/\mathbb{S}_{\bm{x}\bm{x}}^{(k,k)}$ has a finite limiting value.  Hence
\begin{equation}
\label{eq:lindeberg-feller-x2}
\begin{aligned}
\frac{\max_{1\leq i\leq n}\left(x_i^{(k)}\right)^2}{\min(n_1,n_0)\mathbb{S}_{\bm{x}\bm{x}}^{(k,k)}}\rightarrow 0.
\end{aligned}
\end{equation}
Applying Theorem 1 of \cite{li2017general}, we have that, with \eqref{eq:lindeberg-feller-x2}, \\ $\overline{x}_z^{(k)}/\sqrt{(1/n_z-1/n)\mathbb{S}_{\bm{x}\bm{x}}^{(k,k)}}$ converges to a standard normal distribution.
Assumption (2) of Condition \ref{strict_cond} implies that $1/n_z-1/n=o_p(1)$, together with the assumption that $\mathbb{S}_{\bm{x}\bm{x}}^{(k,k)}$ has a finite limiting value (implied by assumption (3) of Condition \ref{strict_cond}), we obtain that $\overline{x}_z^{(k)}=o_p(1)$.

With the fact that $\bm{x}_i$ can be regarded as potential values of the covariates under both $Z_i=0$ and $Z_i=1$,
applying Proposition 3 of \cite{li2017general}, we have $\bm{S}_{\bm{x}\bm{x},z}-\mathbb{S}_{\bm{x}\bm{x}}=o_p(1)$, where $\mathbb{S}_{\bm{x}\bm{x}}$ has a finite non-singular limiting value according to assumption (3) of Condition \ref{strict_cond}. Therefore, applying Slusky's theorem together with the continuous mapping theorem to the right hand side of \eqref{bound_on_projection}, we conclude that $\widehat{\bm{\beta}}_{W,z}^{\top}\overline{\bm{x}}_z=o_p(1)$. In the proof of Theorem~1 we have obtained $\widehat{\tau}_W-p_{co}=o_p(1)$. From (9) and the conclusions above, we know that $\widehat{\tau}_W^{Reg}-p_{co}=o_p(1)+o_p(1)+o_p(1)=o_p(1)$.

Recall that $D_i=B_i-\widehat{\beta}_{B,Z_i}^{\top}\bm{x}_i$.  Hence
\begin{equation*}
\begin{aligned}
\widehat{\tau}_D&=\overline{D}_1-\overline{D}_0\\
&=\overline{B}_1-\overline{B}_0 - \widehat{\bm{\beta}}_{B,1}^{\top}\overline{\bm{x}}_1+\widehat{\bm{\beta}}_{B,0}^{\top}\overline{\bm{x}}_0\\
&=\widehat{\tau}_B- \widehat{\bm{\beta}}_{B,1}^{\top}\overline{\bm{x}}_1+\widehat{\bm{\beta}}_{B,0}^{\top}\overline{\bm{x}}_0.
\end{aligned}
\end{equation*}
Following the proof in Theorem 1, $\widehat{\tau}_B-\tau_B=\widehat{\tau}_B=o_p(1)$.  Following arguments similar as those for proving that $\widehat{\bm{\beta}}_{W,z}^{\top}\overline{\bm{x}}_z=o_p(1)$ for $z=0,1$, we have $\widehat{\bm{\beta}}_{B,z}^{\top}\overline{\bm{x}}_z=o_p(1)$ for $z=0,1$.  Hence $\widehat{\tau}_D=o_p(1)+o_p(1)+o_p(1)=o_p(1)$.

Equations (14) and (11) imply that
\begin{equation*}
	\widehat{\tau}^{Reg}-\tau_{CACE}=\frac{\widehat{\tau}_D}{\widehat{\tau}_W^{Reg}}.
\end{equation*}
Since $p_{co}$ has a positive limit inferior and $\widehat{\tau}_W^{Reg}-p_{co}=o_p(1)$, $\widehat{\tau}_W^{Reg}$ has a positive limit inferior in probability. With $\widehat{\tau}_D=o_p(1)$, we have $\widehat{\tau}_D/\widehat{\tau}_W^{Reg}=o_p(1)$.
Therefore $\widehat{\tau}^{Reg}-\tau_{CACE}=o_p(1)$ as $n\rightarrow\infty$.

\subsection*{S.2.4 Proof of Theorem 4}

Let $\widetilde{\alpha}_{B,z}$ and $\widetilde{\bm{\beta}}_{B,z}$ ($z=0,1$) be the intercept and the vector of coefficients on $\bm{x}_i$ in the linear projection of $B_i(z)$ on $\bm{x}_i$ using all units. That is,
\begin{equation*}
	\label{proj_pop}
	\left(\widetilde{\alpha}_{B,z},\widetilde{\bm{\beta}}_{B,z}\right)=\arg\min_{\alpha,\bm{\beta}}\sum_{i=1}^{n}\left(B_i(z)-\alpha-\bm{\beta}^{\top}\bm{x}_i\right)^2.
\end{equation*}
Define $\widetilde{D}_i=B_i-\widetilde{\bm{\beta}}_{B,Z_i}\bm{x}_i$. Let $\widehat{\tau}_{\widetilde{D}}$ be the difference-in-means for $\widetilde{D}_i$. According to Example 9 in \cite{li2017general}, under Condition \ref{strict_cond}, $\widehat{\tau}_{\widetilde{D}}$ has mean $\tau_B$(=0) and is asymptotically normal, and $\widehat{\tau}_D$ has the same asymptotic distribution with $\widehat{\tau}_{\widetilde{D}}$. Hence, $Var(\widehat{\tau}_{\widetilde{D}})-Var(\widehat{\tau}_{D})=o_p(n^{-1})$ and
\begin{equation*}
	\widehat{\tau}_D/\sqrt{Var(\widehat{\tau}_D)}\stackrel{d}{\rightarrow}N(0,1).
\end{equation*}

\subsection*{S.2.5 Proof of Theorem 5}

We divide the proof into three parts.

{\bf Part 1.} Define $\widehat{D}_i=\widehat{B}_i-\widehat{\bm{\beta}}_{\widehat{B},Z_i}^{\top}\bm{x}_i$. For units with $Z_i=z$ ($z=0,1$), let $S_{\widehat{D}(z)}^2$ be the sample variance of $\widehat{D}_i$. Define
\begin{equation*}
	\label{canonic_hat}
	\widehat{Var}(\widehat{\tau}_D)=\frac{S_{\widehat{D}(1)}^2}{n_1}+\frac{S_{\widehat{D}(0)}^2}{n_0}.
\end{equation*}
We show that \begin{equation}
	\label{confidence_intv_adj}
	\widehat{\tau}^{Reg}\pm \nu_{1-\alpha/2}\sqrt{\widehat{Var}(\widehat{\tau}_D)}/\left|\widehat{\tau}_{W}^{Reg}\right|
\end{equation}
is an asymptotically conservative $(1-\alpha)$ confidence interval for $\tau_{CACE}$.

{\bf Part 2.} We show that
\begin{equation}
		\widehat{Var}^{EHW}(\widehat{\tau}_D)=\widehat{Var}(\widehat{\tau}_D)+o_p(n^{-1}).
\end{equation}
Hence
\begin{equation*}
	\widehat{\tau}^{Reg}\pm \nu_{1-\alpha/2}\sqrt{\widehat{Var}^{EHW}(\widehat{\tau}_D)}/\left|\widehat{\tau}_{W}^{Reg}\right|
\end{equation*}
is an asymptotically conservative $(1-\alpha)$ confidence interval for $\tau_{CACE}$.

{\bf Part 3.} We show that $\widehat{Var}^{HC2}(\widehat{\tau}_D)$ and $\widehat{Var}^{HC3}(\widehat{\tau}_D)$ are both asymptotically equivalent to $\widehat{Var}^{EHW}(\widehat{\tau}_D)$. Therefore
$\widehat{\tau}^{Reg}\pm \nu_{1-\alpha/2}\sqrt{\widehat{Var}^{HC2}(\widehat{\tau}_D)}/\widehat{\tau}_{W}^{Reg}$ and $\widehat{\tau}^{Reg}\pm \nu_{1-\alpha/2}\sqrt{\widehat{Var}^{HC3}(\widehat{\tau}_D)}/\widehat{\tau}_{W}^{Reg}$ are also asymptotically conservative $(1-\alpha)$ confidence intervals for $\tau_{CACE}$.

Details of the proofs are as follows.

{\bf Proof of Part 1:}
Let $\widehat{\bm{\beta}}_{\widehat{B},z}$ denote the vector of coefficients on $\bm{x}_i$ in the linear projection of $\widehat{B}_i$ on $\bm{x}_i$ using units with $Z_i=z$. Recall that $\widehat{B}_i=Y_i-\widehat{\tau}^{Reg}W_i$. Due to the linear relationship between $\widehat{B}_i$ and $(Y_i,W_i)$, we have $\widehat{\bm{\beta}}_{\widehat{B},z}=\widehat{\bm{\beta}}_{Y,z}-\widehat{\tau}^{Reg}\widehat{\bm{\beta}}_{W,z}$. Define $V_i=W_i-\widehat{\beta}_{W,Z_i}^{\top}\bm{x}_i$. For units with $Z_i=z$ ($z=0,1$), let $\overline{V}_z$ be the sample mean of $V_i$, $S_{V(z)}^2$ be the sample variance of $V_i$, $S_{D(z)}^2$ the sample variance of $D_i$, and $S_{D(z),V(z)}$ be the sample covariance between $D_i$ and $V_i$.

Recall that $B_i=Y_i-\tau_{CACE}W_i$, $D_i=B_i-\widehat{\bm{\beta}}_{B,Z_i}^{\top}\bm{x}_i$ and  $\widehat{\bm{\beta}}_{B,z}=\widehat{\bm{\beta}}_{Y,z}-\tau_{CACE}\widehat{\bm{\beta}}_{W,z}$.
For $z=0,1$,
\begin{equation*}
\begin{aligned}
\widehat{D}_i-D_i&=\left(\widehat{B}_i-B_i\right)-\left(\widehat{\bm{\beta}}_{\widehat{B},Z_i}-\widehat{\bm{\beta}}_{B,Z_i}\right)^{\top}\bm{x}_i\\
&=-\left(\widehat{\tau}^{Reg}-\tau_{CACE}\right)W_i+\left(\widehat{\tau}^{Reg}-\tau_{CACE}\right)\widehat{\bm{\beta}}_{W,z}^{\top}\bm{x}_i\\
&=-\left(\widehat{\tau}^{Reg}-\tau_{CACE}\right)V_i.
 \end{aligned}\label{diff_Dhat_D}
 \end{equation*}
Hence we have
\begin{align*} S_{\widehat{D}(z)}^2=S_{D(z)}^2-2\left(\widehat{\tau}^{Reg}-\tau_{CACE}\right)S_{D(z),V(z)}+\left(\widehat{\tau}^{Reg}-\tau_{CACE}\right)^2S_{V(z)}^2.
\end{align*}
The absolute difference between $S_{\widehat{D}(z)}^2$ and $S_{D(z)}^2$ can be bounded by
\begin{equation}
	\begin{aligned}
		&\left|S_{\widehat{D}(z)}^2-S_{D(z)}^2\right|\\
		=&\left|-2\left(\widehat{\tau}^{Reg}-\tau_{CACE}\right)S_{D(z),V(z)}+\left(\tau_{CACE}-\widehat{\tau}^{Reg}\right)^2S_{V(z)}^2\right|\\
		\leq&2\left|\widehat{\tau}^{Reg}-\tau_{CACE}\right|\left|S_{D(z),V(z)}\right|+\left(\widehat{\tau}^{Reg}-\tau_{CACE}\right)^2S_{V(z)}^2.
	\end{aligned}\label{abs_diff_adj}
\end{equation}
Because $V_i$ is the residual of the linear projection of $W_i$ on $\bm{x}_i$, we have
\begin{equation}
	S_{V(z)}^2\leq{}S_{W(z)}^2\leq\frac{n_z}{n_z-1},\label{S_Vz_leq}
\end{equation}
where the second inequality is due to \eqref{S_Wz_leq}.
Because $D_i$ is the residual of the linear projection of $B_i$ on $\bm{x}_i$, we have
\begin{equation}
	S_{D(z)}^2\leq{}S_{B(z)}^2.\label{S_Dz_leq}
\end{equation}
Combining \eqref{S_Vz_leq} and \eqref{S_Dz_leq} with Cauchy-Schwarz inequality, we have
\begin{equation}
	|S_{D(z), V(z)}|\leq\sqrt{S^2_{D(z)}S^2_{V(z)}}\leq\sqrt{\frac{n_z}{n_z-1}S_{B(z)}^2}.\label{S_DzVz_leq}
\end{equation}
Substituting \eqref{S_Vz_leq} and \eqref{S_DzVz_leq} into \eqref{abs_diff_adj}, we obtain the following inequality
\begin{equation}	
	\label{bound_for_variance}
	\begin{aligned}
		&\left|S_{\widehat{D}(z)}^2-S_{D(z)}^2\right|\\
		\leq&2\sqrt{\frac{n_z}{n_z-1}S_{B(z)}^2}\left|\widehat{\tau}^{Reg}-\tau_{CACE}\right|+\frac{n_z}{n_z-1}\left(\widehat{\tau}^{Reg}-\tau_{CACE}\right)^2.
	\end{aligned}
\end{equation}
Applying Proposition 1 of \cite{li2017general}, we have that under Condition \ref{strict_cond},  $S_{B(z)}^2-\mathbb{S}_{B(z)}^2=o_p(1)$, where $\mathbb{S}_{B(z)}^2$ has a finite limiting value. According to Theorem 3, $\widehat{\tau}^{Reg}-\tau_{CACE}=o_p(1)$ as $n\rightarrow\infty$. Applying the continuous mapping theorem, the right hand side of \eqref{bound_for_variance} converges to 0 in probability, which means $S_{\widehat{D}(z)}^2-S_{D(z)}^2=o_p(1)$ as $n\rightarrow\infty$.

Define
\begin{equation*}
	\label{canonic_tilde}
	\widetilde{Var}(\widehat{\tau}_D)=\frac{S_{D(1)}^2}{n_1}+\frac{S_{D(0)}^2}{n_0}.
\end{equation*}
Under assumption (2) of Condition \ref{strict_cond}, $n_1/n$ and $n_0/n$ both have a limit in (0,1).  Hence
$\widehat{Var}(\widehat{\tau}_D)-\widetilde{Var}(\widehat{\tau}_D)=o_p(n^{-1})$. According to Example 9 in \cite{li2017general}, the interval
\begin{equation*}
	\widehat{\tau}_{D}\pm\nu_{1-\alpha/2}\sqrt{\widetilde{Var}(\widehat{\tau}_D)}
\end{equation*}
has at least $1-\alpha$ coverage rate for $\tau_B=0$ as $n\rightarrow\infty$. Hence, the interval
\begin{equation*}
	\widehat{\tau}_{D}\pm\nu_{1-\alpha/2}\sqrt{\widehat{Var}(\widehat{\tau}_D)}
\end{equation*}
also has at least $1-\alpha$ coverage rate for $0$ as $n\rightarrow\infty$.  Following similar arguments to those in Section~4.2, we obtain that an equivalent statement is that
\begin{equation}
	\label{confidence_intv_adj}
	\widehat{\tau}^{Reg}\pm \nu_{1-\alpha/2}\sqrt{\widehat{Var}(\widehat{\tau}_D)}/\left|\widehat{\tau}_{W}^{Reg}\right|
\end{equation}
is an asymptotically conservative $(1-\alpha)$ confidence interval for $\tau_{CACE}$.

{\bf Proof of Part 2:}
Let $\bm{\Omega}$ denote the $n\times(2(K+1))$ design matrix in which the $i$th row is $\bm{\Omega}_i^{\top}=(1,Z_i,\bm{x}_i, Z_i\bm{x}_i)^{\top}$.  Let $\widehat{u}_i$ denote the residual in the OLS regression of $\widehat{B}_i$ on $Z_i$, $\bm{x}_i$ and $Z_i\bm{x}_i$ using all units. The EHW estimator for the covariance matrix of the coefficients is
\begin{equation}
\widehat{\bm{V}}^{EHW}=\left(\sum_{i=1}^n \bm{\Omega}_i\bm{\Omega}_i^{\top}\right)^{-1}\left(\sum_{i=1}^n\widehat{u}_i^2\bm{\Omega}_i\bm{\Omega}_i^{\top}\right)\left(\sum_{i=1}^n \bm{\Omega}_i\bm{\Omega}_i^{\top}\right)^{-1}.\label{eq:V_EHW}
\end{equation}
$\widehat{Var}^{EHW}(\widehat{\tau}_D)$ is the diagonal element of $\widehat{\bm{V}}^{EHW}$ that corresponds to the coefficient for $Z_i$, that is, the second diagonal element of $\widehat{\bm{V}}^{EHW}$.

Denote
\begin{equation}\label{def:G}
\begin{aligned}
	\bm{G}&\equiv n^{-1} \sum_{i=1}^n \bm{\Omega}_i\bm{\Omega}_i^{\top}\\
&=
\left(\begin{array}{cccc}1 &\sum_{i=1}^n Z_i/n &\sum_{i=1}^n \bm{x}_i^{\top}/n & \sum_{i=1}^n Z_i\bm{x}_i^{\top}/n\\
\sum_{i=1}^n Z_i/n & \sum_{i=1}^n Z_i^2/n & \sum_{i=1}^n Z_i\bm{x}_i^{\top}/n & \sum_{i=1}^n Z_i^2\bm{x}_i^{\top}/n\\
\sum_{i=1}^n \bm{x}_i/n & \sum_{i=1}^n Z_i\bm{x}_i/n & \sum_{i=1}^n \bm{x}_i\bm{x}_i^{\top}/n & \sum_{i=1}^n Z_i\bm{x}_i\bm{x}_i^{\top}/n\\
\sum_{i=1}^n Z_i\bm{x}_i/n & \sum_{i=1}^n Z_i^2\bm{x}_i/n & \sum_{i=1}^n Z_i\bm{x}_i\bm{x}_i^{\top}/n & \sum_{i=1}^n Z_i^2\bm{x}_i\bm{x}_i^{\top}/n\end{array}\right)\\
&=\left(\begin{array}{cc}
		\bm{G}_{11} & \bm{G}_{12} \\
		\bm{G}_{21} & \bm{G}_{22}
	\end{array}\right),
\end{aligned}
\end{equation}
where
\begin{equation*}
	\begin{aligned}
		&\bm{G}_{11}=\left(\begin{array}{cc}
			1 & n_1/n \\
			n_1/n & n_1/n
		\end{array}\right),\quad
		\bm{G}_{12}=\left(\begin{array}{cc}
			\bm{0} & n_1/n\cdot\overline{\bm{x}}_1^{\top} \\
			n_1/n\cdot\overline{\bm{x}}_1^{\top} & n_1/n\cdot\overline{\bm{x}}_1^{\top}
		\end{array}\right)\\
		&\bm{G}_{21}=\bm{G}_{12}^{\top},\quad
		\bm{G}_{22}=n^{-1}\sum_{i=1}^{n}\left(\begin{array}{cc}
			\bm{x}_i\bm{x}_i^{\top} & Z_i\bm{x}_i\bm{x}_i^{\top} \\
			Z_i\bm{x}_i\bm{x}_i^{\top} & Z_i\bm{x}_i\bm{x}_i^{\top}
		\end{array}\right),
	\end{aligned}
\end{equation*}
taking into account that $\sum_{i=1}^n Z_i=n_1$, $\sum_{i=1}^n \bm{x}_i/n=\overline{\bm{x}}=\bm{0}$, $\sum_{i=1}^n Z_i\bm{x}_i/n=n_1/n\cdot\sum_{i:\ Z_i=1}\bm{x}_i/n_1=n_1/n\cdot\overline{\bm{x}}_1$ and $Z_i^2=Z_i$.

Denote
\begin{equation*}
	\bm{H}\equiv n^{-2} \sum_{i=1}^n\widehat{u}_i^2\bm{\Omega}_i \bm{\Omega}_i^{\top}=\left(\begin{array}{cc}
		\bm{H}_{11} & \bm{H}_{12} \\
		\bm{H}_{21} & \bm{H}_{22}
	\end{array}\right),
\end{equation*}
where
\begin{equation*}
	\begin{aligned}
		&\bm{H}_{11}=n^{-2}\sum_{i=1}^{n}\left(\begin{array}{cc}
			\widehat{u}_i^2 & Z_i\widehat{u}_i^2 \\
			Z_i\widehat{u}_i^2 & Z_i\widehat{u}_i^2
		\end{array}\right),\quad
		\bm{H}_{12}=n^{-2}\sum_{i=1}^{n}\left(\begin{array}{cc}
			\widehat{u}_i^2\bm{x}_i^{\top} & Z_i\widehat{u}_i^2\bm{x}_i^{\top} \\
			Z_i\widehat{u}_i^2\bm{x}_i^{\top} & Z_i\widehat{u}_i^2\bm{x}_i^{\top}
		\end{array}\right)\\
		&\bm{H}_{21}=\bm{H}_{12}^{\top},\quad
		\bm{H}_{22}=n^{-2}\sum_{i=1}^{n}\left(\begin{array}{cc}
			\widehat{u}_i^2\bm{x}_i\bm{x}_i^{\top} & Z_i\widehat{u}_i^2\bm{x}_i\bm{x}_i^{\top} \\
			Z_i\widehat{u}_i^2\bm{x}_i\bm{x}_i^{\top} & Z_i\widehat{u}_i^2\bm{x}_i\bm{x}_i^{\top}
		\end{array}\right)
	\end{aligned}
\end{equation*}
According to \eqref{eq:V_EHW}, we have
\begin{equation}
\widehat{\bm{V}}^{EHW}=\bm{G}^{-1}\bm{H}\bm{G}^{-1}.\label{eq:V_EHW_short}
\end{equation}

In the proof of Theorem 3, we have shown that $\overline{x}_z^{(k)}/\sqrt{(1/n_z-1/n)\mathbb{S}_{\bm{x}\bm{x}}^{(k,k)}}$ converges to a standard normal distribution for $k=1,\ldots,K$. Since $\mathbb{S}_{\bm{x}\bm{x}}$ has a finite non-singular limit and $n_1/n$ and $n_0/n$ both have a limit in $(0,1)$, we have $\overline{\bm{x}}_z=O_p(n^{-1/2})$ and $\overline{\bm{x}}_z\overline{\bm{x}}_z^{\top}=O_p(n^{-1})$. According to the expression of $\bm{G}_{12}$, we have $\bm{G}_{12}=O_p(n^{-1/2})$. Noting that
\begin{equation*}
\begin{aligned}
\bm{S}_{\bm{x}\bm{x},z}&=\frac{1}{n_z-1}\sum_{i:\ Z_i=z}(\bm{x}_i-\overline{\bm{x}}_z)(\bm{x}_i-\overline{\bm{x}}_z)^{\top}\\
&=\frac{1}{n_z-1}\left(\sum_{i:\ Z_i=z}\bm{x}_i\bm{x}_i^{\top}-n_z\overline{\bm{x}}_z\overline{\bm{x}}_z^{\top}\right),
\end{aligned}
\end{equation*}
and recalling that $\bm{S}_{\bm{x}\bm{x},z}=\mathbb{S}_{\bm{x}\bm{x}}+o_p(1)$ in the proof of Theorem 3, we also have
\begin{equation*}
\begin{aligned}
\sum_{i:\ Z_i=z}\bm{x}_i\bm{x}_i^{\top}/n&=(n_z-1)/n\cdot\bm{S}_{\bm{x}\bm{x},z}+n_z/n\cdot\overline{\bm{x}}_z\overline{\bm{x}}_z^{\top}\\
&=(n_z-1)/n\cdot\mathbb{S}_{\bm{x}\bm{x}}+o_p(1)+O_p(n^{-1})\\
&=(n_z-1)/n\cdot\mathbb{S}_{\bm{x}\bm{x}}+o_p(1).
\end{aligned}
\end{equation*}
Hence, $\sum_{i=1}^n \bm{x}_i\bm{x}_i^{\top}/n=\sum_{i:\ Z_i=1}\bm{x}_i\bm{x}_i^{\top}/n+\sum_{i:\ Z_i=0}\bm{x}_i\bm{x}_i^{\top}/n$ converges in probability to a non-singular matrix, and $\sum_{i=1}^n Z_i\bm{x}_i\bm{x}_i^{\top}/n=\sum_{i:\ Z_i=1}\bm{x}_i\bm{x}_i^{\top}/n$ also converges in probability to a non-singular matrix.  Using the expression of $\bm{G}_{22}$, we obtain that $\bm{G}_{22}$ converges in probability to a non-singular matrix. Since $n_1/n$ has a limit in $(0,1)$, $\bm{G}_{11}$ also converges to a non-singular matrix.  Overall, $\bm{G}$ converges in probability to a non-singular matrix.

Let
\begin{equation*}
	\bm{G}_{diag}=\left(\begin{array}{cc}
		\bm{G}_{11} & \bm{0} \\
		\bm{0} & \bm{G}_{22}
	\end{array}\right),\quad
	\bm{G}_{diff}=\left(\begin{array}{cc}
		\bm{0} & \bm{G}_{12} \\
		\bm{G}_{21} & \bm{0}
	\end{array}\right).
\end{equation*}
We have $\bm{G}=\bm{G}_{diag}+\bm{G}_{diff}$.  According to previous results, $\bm{G}_{diff}=O_p(n^{-1/2})$ and $\bm{G}_{diag}$ converges in probability to a non-singular matrix. According to Hua's identity, if two square matrices $\bm{M}$ and $\bm{N}$ are non-singular and $\bm{M}-\bm{N}^{-1}$ is also non-singular, then $(\bm{M}-\bm{N}^{-1})^{-1}-\bm{M}^{-1}$ is non-singular, with $(\bm{M}-\bm{N}^{-1})^{-1}-\bm{M}^{-1}=(\bm{M}\bm{N}\bm{M}-\bm{M})^{-1}$.  Plugging in $\bm{M}=\bm{G}_{diag}$ and $\bm{N}=-\bm{G}_{diff}^{-1}$, we have
\begin{equation}	\bm{G}_{diag}^{-1}-\bm{G}^{-1}=(\bm{G}_{diag}+\bm{G}_{diag}\bm{G}_{diff}^{-1}\bm{G}_{diag})^{-1}=O_p(n^{-1/2}).\label{eq:Ginv_diff}
\end{equation}

Consider the linear projection of $\widehat{B}_i$ on $Z_i$, $\bm{x}_i$ and $Z_i\bm{x}_i$.
Let $\widehat{\beta}_0^*$, $\widehat{\beta}_Z^*$, $\widehat{\bm{\beta}}_{\bm{x}}^*$ and $\widehat{\bm{\beta}}_{Z\bm{x}}^*$ denote respectively the intercept, the coefficient on $Z_i$, the vector of coefficients on $\bm{x}_i$ and the vector of coefficients on $Z_i\bm{x}_i$. Recall that $\widehat{\bm{\beta}}_{\widehat{B},z}$ is the vector of coefficients on $\bm{x}_i$ in the linear projection of $\widehat{B}_i$ on $\bm{x}_i$ using units with $Z_i=z$ for $z=0,1$.  According to the proof of Proposition A2, we have
$\widehat{\bm{\beta}}_{\widehat{B},0}=\widehat{\bm{\beta}}_{\bm{x}}^*$, $\widehat{\bm{\beta}}_{\widehat{B},1}=\widehat{\bm{\beta}}_{\bm{x}}^*+\widehat{\bm{\beta}}_{Z\bm{x}}^*$, $\sum_{i:\ Z_i=0}\widehat{u}_i=\sum_{i:\ Z_i=1}\widehat{u}_i=0$. Recall that $\widehat{D}_i=\widehat{B}_i-\widehat{\bm{\beta}}_{\widehat{B},Z_i}^{\top}\bm{x}_i$.  According to the definition of $\widehat{u}_i$, we have that for units with $Z_i=0$, $\widehat{u}_i=\widehat{B}_i-\widehat{\beta}_0^*-(\widehat{\bm{\beta}}_{\bm{x}}^*)^{\top}\bm{x}_i=\widehat{D}_i-\widehat{\beta}_0^*$, and that for units with $Z_i=1$, $\widehat{u}_i=\widehat{B}_i-\widehat{\beta}_0^*-\widehat{\beta}_Z^*-(\widehat{\bm{\beta}}_{\bm{x}}^*)^{\top}\bm{x}_i-(\widehat{\bm{\beta}}_{Z\bm{x}}^*)^{\top}\bm{x}_i=\widehat{D}_i-\widehat{\beta}_0^*-\widehat{\beta}_Z^*$. Hence,
\begin{equation*}
\begin{aligned}
\overline{\widehat{D}}_0&\equiv \frac{1}{n_0}\sum_{i:\ Z_i=0}\widehat{D}_i\\
&=\frac{1}{n_0}\sum_{i:\ Z_i=0}\left(\widehat{u}_i+\widehat{\beta}_0^*\right)=\widehat{\beta}_0^*,\\
\overline{\widehat{D}}_1&\equiv \frac{1}{n_1}\sum_{i:\ Z_i=1}\widehat{D}_i\\
&=\frac{1}{n_1}\sum_{i:\ Z_i=1}\left(\widehat{u}_i+\widehat{\beta}_0^*+\widehat{\beta}_Z^*\right)=\widehat{\beta}_0^*+\widehat{\beta}_Z^*,\\
S_{\widehat{D}(0)}^2&\equiv\frac{1}{n_0-1}\sum_{i:\ Z_i=0}\left(\widehat{D}_i-\overline{\widehat{D}}_0\right)^2\\
&=\frac{1}{n_0-1}\sum_{i:\ Z_i=0}\left(\widehat{u}_i+\widehat{\beta}_0^*-\widehat{\beta}_0^*\right)^2=\frac{1}{n_0-1}\sum_{i:\ Z_i=0}\widehat{u}_i^2,\\
S_{\widehat{D}(1)}^2&\equiv\frac{1}{n_1-1}\sum_{i:\ Z_i=1}\left(\widehat{D}_i-\overline{\widehat{D}}_0\right)^2\\
&=\frac{1}{n_1-1}\sum_{i:\ Z_i=1}\left(\widehat{u}_i+\widehat{\beta}_0^*+\widehat{\beta}_Z^*-\widehat{\beta}_0^*-\widehat{\beta}_Z^*\right)^2=\frac{1}{n_1-1}\sum_{i:\ Z_i=1}\widehat{u}_i^2.
\end{aligned}
\end{equation*}
Therefore we have for $z=0,1$,
\begin{equation}
n_z^{-1}\sum_{i:\ Z_i=z}\widehat{u}_i^2=n_z^{-1}(n_z-1)S^2_{\widehat{D}(z)}. \label{eq:hatu2}
\end{equation}

For $z=0,1$, recall that $S^2_{\widehat{D}(z)}-S^2_{D(z)}=o_p(1)$. According to Example 9 of \cite{li2017general}, $S^2_{D(z)}-\mathbb{S}^2_{\widetilde{D}(z)}=o_p(1)$, where $\mathbb{S}^2_{\widetilde{D}(z)}$ is the population variance of $\widetilde{D}_i(z)=B_i(z)-\widetilde{\bm{\beta}}_{B,z}\bm{x}_i$ for $z=0,1$ ($\widetilde{\bm{\beta}}_{B,z}$ ($z=0,1$) is defined at the beginning of the proof of Theorem 4 to be the vector of coefficients on $\bm{x}_i$ in the linear projection of $B_i(z)$ on $\bm{x}_i$ using all units). Combining these results with \eqref{eq:hatu2}, we have
\begin{equation*}
	n_z^{-1}\sum_{i:\ Z_i=z}\widehat{u}_i^2=S^2_{D(z)}+o_p(1)=\mathbb{S}^2_{\widetilde{D}(z)}+o_p(1),\quad{}z=0,1.
\end{equation*}
Since $\widetilde{D}_i(z)$ is the projection residual of $B_i(z)$, we have $\mathbb{S}^2_{\widetilde{D}(z)}\leq\mathbb{S}^2_{B(z)}$, with $\mathbb{S}^2_{B(z)}$ having a finite limit for $z=0,1$.
Under assumption (2) of Condition \ref{strict_cond}, $n_1/n$ and $n_0/n$ both have a limit in $(0,1)$.  Thus, $\sum_{i:\ Z_i=z}\widehat{u}_i^2=O_p(n)$ for $z=0,1$. It follows that $\sum_{i=1}^n\widehat{u}_i^2=\sum_{i:\ Z_i=1}\widehat{u}_i^2+\sum_{i:\ Z_i=0}\widehat{u}_i^2=O_p(n)$, and $\sum_{i=1}^n Z_i\widehat{u}_i^2=\sum_{i:\ Z_i=1}\widehat{u}_i^2=O_p(n)$.  According to the definition of $\bm{H}_{11}$, we have $\bm{H}_{11}=O_p(n^{-1})$.

Assumption (4) of Condition \ref{strict_cond} implies that $\max_{1\leq{}i\leq{}n}|x_i^{(k)}|=o_p(n^{1/2})$.  According to the definition of $\bm{H}_{12}$, each element of $\bm{H}_{12}$ is either in the form of $\sum_{i=1}^n \widehat{u}_i^2x_i^{(k)}/n^2=\sum_{i:\ Z_i=1} \widehat{u}_i^2x_i^{(k)}/n^2+\sum_{i:\ Z_i=0} \widehat{u}_i^2x_i^{(k)}/n^2$, or in the form of $\sum_{i=1}^n Z_i\widehat{u}_i^2x_i^{(k)}/n^2=\sum_{i:\ Z_i=1} \widehat{u}_i^2x_i^{(k)}/n^2$.  For $z=0,1$, we have
\begin{equation*}
\begin{aligned}
&\left|\frac{1}{n^2}\sum_{i:\ Z_i=z}\widehat{u}_i^2x_i^{(k)}\right|\\
\leq & \frac{1}{n^2}\left(\sum_{i:\ Z_i=z} \widehat{u}_i^2\right)\left(\max_{1\leq{}i\leq{}n}|x_i^{(k)}|\right)\\
=&\frac{1}{n^2}\cdot O_p(n)\cdot o_p(n^{1/2})=o_p(n^{-1/2}).
\end{aligned}
\end{equation*}
Hence $\sum_{i=1}^n \widehat{u}_i^2x_i^{(k)}/n^2=o_p(n^{-1/2})$, and $\sum_{i=1}^n Z_i\widehat{u}_i^2x_i^{(k)}/n^2=o_p(n^{-1/2})$.  Therefore, $\bm{H}_{12}=o_p(n^{-1/2})$.

According to the definition of $\bm{H}_{22}$, each element of $\bm{H}_{22}$ is either in the form of $\sum_{i=1}^n \widehat{u}_i^2x_i^{(k)}x_i^{(l)}/n^2=\sum_{i:\ Z_i=1} \widehat{u}_i^2x_i^{(k)}x_i^{(l)}/n^2+\sum_{i:\ Z_i=0} \widehat{u}_i^2x_i^{(k)}x_i^{(l)}/n^2$, or in the form of \\$\sum_{i=1}^n Z_i\widehat{u}_i^2x_i^{(k)}x_i^{(l)}/n^2=\sum_{i:\ Z_i=1} \widehat{u}_i^2x_i^{(k)}x_i^{(l)}/n^2$.  For $z=0,1$, we have
\begin{equation*}
\begin{aligned}
&\left|\frac{1}{n^2}\sum_{i:\ Z_i=z}\widehat{u}_i^2x_i^{(k)}x_i^{(l)}\right|\\
\leq & \frac{1}{n^2}\left(\sum_{i:\ Z_i=z} \widehat{u}_i^2\right)\left(\max_{1\leq{}i\leq{}n}|x_i^{(k)}|\right)\left(\max_{1\leq{}i\leq{}n}|x_i^{(l)}|\right)\\
=&\frac{1}{n^2}\cdot O_p(n)\cdot o_p(n^{1/2})\cdot o_p(n^{1/2})=o_p(1).
\end{aligned}
\end{equation*}
Hence $\sum_{i=1}^n \widehat{u}_i^2x_i^{(k)}x_i^{(l)}/n^2=o_p(1)$, and $\sum_{i=1}^n Z_i\widehat{u}_i^2x_i^{(k)}x_i^{(l)}/n^2=o_p(1)$.  Therefore, $\bm{H}_{22}=o_p(1)$.  In summary, we have $\bm{H}=o_p(1)$.

With \eqref{eq:V_EHW_short}, we have
\begin{equation*}
	\begin{aligned}
		&\widehat{\bm{V}}^{EHW}-\bm{G}_{diag}^{-1}\bm{H}\bm{G}_{diag}^{-1}\\
		=&\bm{G}^{-1}\bm{H}\bm{G}^{-1}-\bm{G}_{diag}^{-1}\bm{H}\bm{G}_{diag}^{-1}\\
		=&(\bm{G}^{-1}-\bm{G}_{diag}^{-1})\bm{H}\bm{G}^{-1}+\bm{G}_{diag}^{-1}\bm{H}(\bm{G}^{-1}-\bm{G}_{diag}^{-1})\\
		=&(\bm{G}^{-1}-\bm{G}_{diag}^{-1})\bm{H}\bm{G}_{diag}^{-1}+\bm{G}_{diag}^{-1}\bm{H}(\bm{G}^{-1}-\bm{G}_{diag}^{-1})\\
		&+(\bm{G}^{-1}-\bm{G}_{diag}^{-1})\bm{H}(\bm{G}^{-1}-\bm{G}_{diag}^{-1})
\end{aligned}
\end{equation*}
According to \eqref{eq:Ginv_diff}, we have $\bm{G}^{-1}-\bm{G}_{diag}^{-1}=O_p(n^{-1/2})$.  We then have $(\bm{G}^{-1}-\bm{G}_{diag}^{-1})\bm{H}(\bm{G}^{-1}-\bm{G}_{diag}^{-1})=O_p(n^{-1/2})\cdot o_p(1) \cdot O_p(n^{-1/2})=o_p(n^{-1})$. Since $\bm{G}_{diag}$ is a block-diagonal matrix, we have
\begin{equation*}
\bm{G}_{diag}^{-1}=\left(\begin{array}{cc}
			\bm{G}_{11}^{-1} & \bm{0} \\
			\bm{0} & \bm{G}_{22}^{-1}
		\end{array}\right).
\end{equation*}
Hence, we obtain that
\begin{equation}\label{eq:V_EHW_diff}
	\begin{aligned}
		\widehat{\bm{V}}^{EHW}-\bm{G}_{diag}^{-1}\bm{H}\bm{G}_{diag}^{-1}=&O_p(n^{-1/2})\left(\begin{array}{cc}
			\bm{H}_{11}\bm{G}_{11}^{-1} & \bm{H}_{12}\bm{G}_{22}^{-1} \\
			\bm{H}_{21}\bm{G}_{11}^{-1} & \bm{H}_{22}\bm{G}_{22}^{-1}
		\end{array}\right)\\
		&+\left(\begin{array}{cc}
			\bm{G}_{11}^{-1}\bm{H}_{11} & \bm{G}_{11}^{-1}\bm{H}_{12} \\
			\bm{G}_{22}^{-1}\bm{H}_{21} & \bm{G}_{22}^{-1}\bm{H}_{22}
		\end{array}\right)O_p(n^{-1/2})+o_p(n^{-1})
	\end{aligned}
\end{equation}

Let $\widehat{\bm{V}}^{EHW}_{[2,2]}$ denote the submatrix of $\widehat{\bm{V}}^{EHW}$ that consists of the first two rows and the first two columns. It is easy to see that the corresponding submatrix of $\bm{G}_{diag}^{-1}\bm{H}\bm{G}_{diag}^{-1}$ equals $\bm{G}_{11}^{-1}\bm{H}_{11}\bm{G}_{11}^{-1}$. Taking the submatrix of \eqref{eq:V_EHW_diff}, we have
\begin{equation*}
\begin{aligned}
\widehat{\bm{V}}^{EHW}_{[2,2]}-\bm{G}_{11}^{-1}\bm{H}_{11}\bm{G}_{11}^{-1}=O_p(n^{-1/2})\cdot \bm{H}_{11}\bm{G}_{11}^{-1}+\bm{G}_{11}^{-1}\bm{H}_{11}\cdot O_p(n^{-1/2}) + o_p(n^{-1})
\end{aligned}
\end{equation*}
Since $\bm{G}_{11}$ converges in probability to a non-singular matrix and $\bm{H}_{11}=O_p(n^{-1})$, we obtain that $\widehat{\bm{V}}^{EHW}_{[2,2]}-\bm{G}_{11}^{-1}\bm{H}_{11}\bm{G}_{11}^{-1}=o_p(n^{-1})$.

Plugging in the expressions for $\bm{G}_{11}$ and $\bm{H}_{11}$, we have
\begin{align*}
	&\bm{G}_{11}^{-1}\bm{H}_{11}\bm{G}_{11}^{-1}\\=&\left(\begin{array}{cc}
		1/n_0 & -1/n_0 \\
		-1/n_0 & n/(n_1n_0)
	\end{array}\right)
	\sum_{i=1}^{n}\left(\begin{array}{cc}
		\widehat{u}_i^2 & Z_i\widehat{u}_i^2 \\
		Z_i\widehat{u}_i^2 & Z_i\widehat{u}_i^2
	\end{array}\right)
	\left(\begin{array}{cc}
		1/n_0 & -1/n_0 \\
		-1/n_0 & n/(n_1n_0)
	\end{array}\right)\\
	=&\left(\begin{array}{cc}
		n_0^{-2}\sum_{i:Z_{i}=0}\widehat{u}_i^2 & -n_0^{-2}\sum_{i:Z_{i}=0}\widehat{u}_i^2 \\
		-n_0^{-2}\sum_{i:Z_{i}=0}\widehat{u}_i^2 & n_1^{-2}\sum_{i:Z_{i}=1}\widehat{u}_i^2+n_0^{-2}\sum_{i:Z_{i}=0}\widehat{u}_i^2
	\end{array}\right).
\end{align*}
By definition, $\widehat{Var}^{EHW}(\widehat{\tau}_D)$ is the element of $\widehat{\bm{V}}^{EHW}_{[2,2]}$ that is in the second row and second column. Hence we have
\begin{equation} \widehat{Var}^{EHW}(\widehat{\tau}_D)=\frac{1}{n_1^2}\sum_{i:Z_{i}=1}\widehat{u}_i^2+\frac{1}{n_0^2}\sum_{i:Z_{i}=0}\widehat{u}_i^2+o_p(n^{-1}).\label{eq:varhat_EHW_Dhat}
\end{equation}

By definition, $\widehat{Var}(\widehat{\tau}_D)=S_{\widehat{D}(1)}^2/n_1+S_{\widehat{D}(0)}^2/n_0$.  Combining this with \eqref{eq:hatu2}, we have
\begin{equation} \widehat{Var}(\widehat{\tau}_D)=\frac{\sum_{i:Z_{i}=1}\widehat{u}_i^2}{n_1(n_1-1)}+\frac{\sum_{i:Z_{i}=0}\widehat{u}_i^2}{n_0(n_0-1)}.
\end{equation}
Combining this with \eqref{eq:varhat_EHW_Dhat}, we have
\begin{equation}
		\widehat{Var}^{EHW}(\widehat{\tau}_D) =\widehat{Var}(\widehat{\tau}_D)-\frac{\sum_{i:Z_{i}=1}\widehat{u}_i^2}{n_1^2(n_1-1)}-\frac{\sum_{i:Z_{i}=0}\widehat{u}_i^2}{n_0^2(n_0-1)}+o_p(n^{-1}).
\end{equation}
Recall that $\sum_{Z_i=z}\widehat{u}_i^2=O_p(n)$ for $z=0,1$.  Also $n_1/n$ and $n_0/n$ both have a limit in $(0,1)$.  Hence we have
\begin{equation*}
\widehat{Var}^{EHW}(\widehat{\tau}_D)=\widehat{Var}(\widehat{\tau}_D)+o_p(n^{-1}).
\end{equation*}
Combining this with the statement that \eqref{confidence_intv_adj} is an asymptotically conservative $(1-\alpha)$ confidence interval for $\tau_{CACE}$, we obtain that
\begin{equation*}
	\widehat{\tau}^{Reg}\pm \nu_{1-\alpha/2}\sqrt{\widehat{Var}^{EHW}(\widehat{\tau}_D)}/\left|\widehat{\tau}_{W}^{Reg}\right|
\end{equation*}
is an asymptotically conservative $(1-\alpha)$ confidence interval for $\tau_{CACE}$.

{\bf Proof of Part 3:}
Let $h_i$ denote the $i$th diagonal element of $\bm{\Omega}(\bm{\Omega}^{\top}\bm{\Omega})^{-1}\bm{\Omega}^{\top}$.
The HC2 estimator for the covariance matrix of the coefficients replaces $\widehat{u}_i$ with $\widehat{u}_i/\sqrt{1-h_i}$, and the HC3 estimator for the covariance matrix of the coefficients replaces $\widehat{u}_i$ with $\widehat{u}_i/(1-h_i)$. According to the definition of $\bm{G}$ in \eqref{def:G}, we have $\bm{\Omega}^{\top}\bm{\Omega}=n\bm{G}$.  Previously we have proved that $\bm{G}$ converges in probability to a non-singular matrix.  Hence $(\bm{\Omega}^{\top}\bm{\Omega})^{-1}=O_p(n^{-1})$. Also, assumption (4) of Condition \ref{strict_cond} implies that $\bm{x}_i=o(n^{1/2})$, and hence $\bm{\Omega}=o(n^{1/2})$. Therefore, $h_i=o(n^{1/2})\cdot{}O_p(n^{-1})\cdot{}o(n^{1/2})=o_p(1)$. It follows that the scale multiplier $1/\sqrt{1-h_i}$ in the HC2 estimator and the scale multiplier $1/(1-h_i)$ in the HC3 estimator both converge to 1 as $n\rightarrow\infty$. Hence, $\widehat{Var}^{HC2}(\widehat{\tau}_D)$ and $\widehat{Var}^{HC3}(\widehat{\tau}_D)$ are both asymptotically equivalent to $\widehat{Var}^{EHW}(\widehat{\tau}_D)$.

\subsection*{S.2.6 Proof of Theorem 6}
In the proof of Theorem~2 (the last paragraph in that proof), we have shown $\widehat{Var}(\widehat{\tau}_B)/Var(\widehat{\tau}_B)^{+}$ $\stackrel{p}{\longrightarrow}1$ as $n\rightarrow\infty$. Under assumptions (2) and (3) of Condition \ref{strict_cond}, $n\cdot{}Var(\widehat{\tau}_B)^{+}$ has a finite limiting value as $n\rightarrow\infty$, implying that $Var(\widehat{\tau}_B)^{+}=O_p(n^{-1})$ and $\widehat{Var}(\widehat{\tau}_B)-Var(\widehat{\tau}_B)^{+}=o_p(n^{-1})$. Furthermore,
\begin{equation*}
	\begin{aligned} &\left|\frac{\sqrt{\widehat{Var}(\widehat{\tau}_B)}}{|\widehat{\tau}_W|}-\frac{\sqrt{Var(\widehat{\tau}_B)^{+}}}{p_{co}}\right|\\
	=&\left|\frac{\sqrt{\widehat{Var}(\widehat{\tau}_B)}-\sqrt{Var(\widehat{\tau}_B)^{+}}}{|\widehat{\tau}_W|}-\frac{\sqrt{Var(\widehat{\tau}_B)^{+}}(|\widehat{\tau}_W|-p_{co})}{|\widehat{\tau}_W|{}p_{co}}\right|\\
	\leq&\left|\frac{\sqrt{\widehat{Var}(\widehat{\tau}_B)}-\sqrt{Var(\widehat{\tau}_B)^{+}}}{|\widehat{\tau}_W|}\right|+\left|\frac{\sqrt{Var(\widehat{\tau}_B)^{+}}(|\widehat{\tau}_W|-p_{co})}{|\widehat{\tau}_W|{}p_{co}}\right|.
	\end{aligned}
\end{equation*}
We also know from the proof of Theorem 1 that $\widehat{\tau}_W-p_{co}=o_p(1)$.
Under assumption (1) of Condition \ref{strict_cond}, $\widehat{\tau}_W$ and $\widehat{\tau}_Wp_{co}$ have positive limit inferiors in probability as $n\rightarrow\infty$. Hence $|\widehat{\tau}_W|-p_{co}=o_p(1)$, and $|\widehat{\tau}_W|$ and $|\widehat{\tau}_W|p_{co}$ have positive limit inferiors in probability as $n\rightarrow\infty$. It follows from Slutsky's theorem and the continous mapping theorem that
\begin{equation}
\sqrt{\widehat{Var}(\widehat{\tau}_B)}/|\widehat{\tau}_W|-\sqrt{Var(\widehat{\tau}_B)^{+}}/p_{co}=o_p(n^{-1/2}).\label{eq:divide_by_tauWhat_pco}
\end{equation}

For $z=0,1$, let $\mathbb{S}^2_{\widetilde{D}(z)}$ be the finite population variance of $\widetilde{D}_i(z)$ in treatment arm $z$. As $\widetilde{D}_i(z)$ is the projection residual of $B_i(z)$, we have
\begin{equation}
\label{eq:var_tildeDz}
\begin{aligned}
\mathbb{S}^2_{\widetilde{D}(z)}&=\mathbb{S}^2_{B(z)}-\mathbb{S}^2_{B(z)\mid\bm{x}}\\
&=\mathbb{S}^2_{B(z)}-\mathbb{S}_{B(z),\bm{x}}\mathbb{S}_{\bm{x}\bm{x}}^{-1}\mathbb{S}_{B(z),\bm{x}}^{\top}.
\end{aligned}
\end{equation}
(Recall that $\mathbb{S}^2_{B(z)}$ is the finite population variance of $B_i(z)$, $\mathbb{S}^2_{B(z)\mid\bm{x}}$ is the finite
population variance of the linear projection of $B_i(z)$ on $\bm{x}_i$, $\mathbb{S}_{B(z),\bm{x}}$ is the finite population covariance vector between $B_i(z)$ and $\bm{x}_i$, and $\mathbb{S}_{\bm{x}\bm{x}}$ is the finite population covariance matrix for $\bm{x}_i$.) It follows from assumption (3) of Condition \ref{strict_cond} that $\mathbb{S}^2_{\widetilde{D}(z)}$ has a finite limiting value.

According to Example 9 of \cite{li2017general}, $S^2_{D(z)}-\mathbb{S}^2_{\widetilde{D}(z)}=o_p(1)$. It is also shown in the proof of Theorem 5 that $S^2_{\widehat{D}(z)}-S^2_{D(z)}=o_p(1)$. Hence $S^2_{\widehat{D}(z)}-\mathbb{S}^2_{\widetilde{D}(z)}=o_p(1)$. By definition, $\widehat{Var}(\widehat{\tau}_D)=S_{\widehat{D}(1)}^2/n_1+S_{\widehat{D}(0)}^2/n_0$. As $n_1/n$ and $n_0/n$ have finite limits in $(0,1)$ according to assumption (2) of Condition \ref{strict_cond}, we have
\begin{equation*} \widehat{Var}(\widehat{\tau}_D)-\left(\frac{\mathbb{S}^2_{\widetilde{D}(1)}}{n_1}+\frac{\mathbb{S}^2_{\widetilde{D}(0)}}{n_0}\right)=o_p(n^{-1}).
\end{equation*}
Combining the first line of \eqref{eq:var_tildeDz} with (18) and (19), we have
\begin{equation*}
\frac{\mathbb{S}^2_{\widetilde{D}(1)}}{n_1}+\frac{\mathbb{S}^2_{\widetilde{D}(0)}}{n_0}=Var\left(\widehat{\tau}_B\right)^{+}-Var(\widehat{\tau}_{B\mid\bm{x}})^{+}.
\end{equation*}
Hence
\begin{equation*} \widehat{Var}(\widehat{\tau}_D)-\left(Var\left(\widehat{\tau}_B\right)^{+}-Var(\widehat{\tau}_{B\mid\bm{x}})^{+}\right)=o_p(n^{-1}).
\end{equation*}
Following similar arguments that lead to \eqref{eq:divide_by_tauWhat_pco}, we have
\begin{equation} \sqrt{\widehat{Var}(\widehat{\tau}_D)}/|\widehat{\tau}_W^{Reg}|-\sqrt{Var\left(\widehat{\tau}_B\right)^{+}-Var(\widehat{\tau}_{B\mid\bm{x}})^{+}}/p_{co}=o_p(n^{-1/2}). \label{eq:divide2_by_tauWhat_pco}
\end{equation}

With \eqref{eq:divide_by_tauWhat_pco} and \eqref{eq:divide2_by_tauWhat_pco}, we have
\begin{equation}
	\label{PRIL_sample}
	\begin{aligned}		&1-\frac{2\sqrt{\widehat{Var}(\widehat{\tau}_D)}/|\widehat{\tau}_W^{Reg}|}{2\sqrt{\widehat{Var}(\widehat{\tau}_B)}/|\widehat{\tau}_W|}\\		=&1-\frac{\sqrt{Var\left(\widehat{\tau}_B\right)^{+}-Var(\widehat{\tau}_{B\mid\bm{x}})^{+}}/p_{co}+o_p(n^{-1/2})}{\sqrt{Var(\widehat{\tau}_B)^{+}}/p_{co}+o_p(n^{-1/2})}\\
		=&1-\sqrt{1-Var(\widehat{\tau}_{B\mid\bm{x}})^{+}/Var(\widehat{\tau}_B)^{+}}+o_p(1)\quad\text{as }n\rightarrow\infty.
	\end{aligned}
\end{equation}
By definition, the PRIL of (15) compared with (8) is
\begin{equation*}
1-\frac{2\sqrt{\widehat{Var}^{EHW}(\widehat{\tau}_D)}/|\widehat{\tau}_W^{Reg}|}
{2\sqrt{\widehat{Var}(\widehat{\tau}_B)}/|\widehat{\tau}_W|}
\end{equation*}
In the proof of Theorem 5, it has been shown that $\widehat{Var}^{EHW}(\widehat{\tau}_D)$ is asymptotically equivalent to $\widehat{Var}(\widehat{\tau}_D)$.  Combining this with \eqref{PRIL_sample}, we obtain that the PRIL of (15) has a limit $1-\sqrt{1-Var(\widehat{\tau}_{B\mid\bm{x}})^{+}/Var(\widehat{\tau}_B)^{+}}$.  Similarly, the PRIL of (16) or (17) also has a limit $1-\sqrt{1-Var(\widehat{\tau}_{B\mid\bm{x}})^{+}/Var(\widehat{\tau}_B)^{+}}$.

\bibliographystyle{apalike}
\bibliography{Refs}
\end{document}